\documentclass[10pt,journal,compsoc]{IEEEtran}

\usepackage{times}
\usepackage{amssymb}
\usepackage{times,amsmath,epsfig,subfigure}
\usepackage{algorithm}
\usepackage[noend]{algorithmic}
\usepackage{graphicx}
\usepackage{color}
\usepackage{balance}

\usepackage{multirow}
\newtheorem{definition}{Definition}

\newcommand{\comment}[1]{}

\ifCLASSOPTIONcompsoc
  \usepackage[nocompress]{cite}
\else
  \usepackage{cite}
\fi
\ifCLASSINFOpdf
\else
\fi
\begin{document}
%

\title{Homogeneous and Mixed Energy Communities Discovery with Spatial-Temporal Net Energy}

%
%
%
%

\author{Shangyu~Xie,
        Han~Wang,
        Shengbin~Wang, Haibing~Lu, Yuan~Hong, Dong~Jin
        and~Qi~Liu
\IEEEcompsocitemizethanks{\IEEEcompsocthanksitem S.~Xie, H.~Wang, Y.~Hong and D.~Jin are with the Department of Computer Science, Illinois Institute of Technology, Chicago, IL.
\IEEEcompsocthanksitem S. Wang is with University of Wisconsin at Eau Claire.
\IEEEcompsocthanksitem H. Lu is with Santa Clara University.
\IEEEcompsocthanksitem Q. Liu is with University of Rhode Island.

Corresponding E-mail: yuan.hong@iit.edu
}

}

\IEEEtitleabstractindextext{%
\begin{abstract}
Smart grid has integrated an increasing number of distributed energy resources to improve the efficiency and flexibility of power generation and consumption as well as the resilience of the power grid. The energy consumers on the power grid (e.g., households) equipped with the distributed energy resources can be considered as ``microgrids'' that both generate and consume electricity. In this paper, we study the energy community discovery problems which identify multiple kinds of energy communities for the microgrids to facilitate energy management (e.g., power supply adjustment, load balancing, energy sharing) on the grid, such as homogeneous energy communities (HECs), mixed energy communities (MECs), and self-sufficient energy communities (SECs). Specifically, we present efficient algorithms to discover such communities of microgrids by taking into account not only their geo-locations but also their net energy over any period. Finally, we experimentally validate the performance of the algorithms using both synthetic and real datasets.
\end{abstract}

}

\maketitle

\IEEEdisplaynontitleabstractindextext

%
\IEEEpeerreviewmaketitle

\section{Introduction}
\label{sec:intro}Smart grid superposes a communication network on top of the electrical power network allowing massive sensor data collection from the grid as well as two-way metering of power for users \cite{FangMXY12}. While collecting and transmitting data across the grid, it allows for the integration of renewable energy resources at the individual consumer level \cite{sgsg}. It creates a paradigm where any individual consumer on the grid can also be a supplier of power: this facilitates the creation of microgrids. Microgrids are localized grids that can be separated from the larger power grid to operate autonomously and be self-sufficient in power. A microgrid typically consists of renewable (wind turbines, solar panels, etc.) and/or non-renewable (micro-turbines, fuel cells, etc.) energy resources, energy storage devices, and energy consuming devices/appliances, all of which are connected through a power and communication network \cite{NationalGridWorkshop}. A microgrid can be operated in a grid with the connected or islanded mode. In the islanded mode, it could be connected to other microgrids or operate independently. Therefore, microgrids can provide energy independence to individual communities or entities who intend to manage their own power generation and distribution \cite{MG02}. Moreover, microgrids can provide resilience against large-scale failures across the grid: they can continue to operate if large-scale blackouts occur \cite{MG02}.

With autonomous energy, every microgrid may fully or partially feed their local demand. More importantly, numerous microgrids would have great flexibility to utilize their local energy to collaboratively advance the energy management on the power grid, e.g., load balancing \cite{loadmanagement}, energy exchange/sharing \cite{SaadHPB12}, and load shifting \cite{LiangLLLS13}. Therefore, it is desirable to discover various microgrid communities that can efficiently implement their cooperation on the grid \cite{CommunityEnergy,ZeroEnergy}. For instance, 

First, the grid can identify communities for a given set of microgrids, all of which need external power supply, and then the grid can adjust the power supply to different communities by placing new generators at different locations or substations\footnotemark[1]. On the contrary, the grid can identify communities for a given set of microgrids, all of which have excessive electricity, and then the grid can also adjust the power supply by reducing it to such communities, or by establishing energy banks to store excessive energy at different locations. 
	
\footnotetext[1]{New routes for transmitting electricity to each community can also be established if it complies with the development strategy of the power grid.}
	
Second, the grid can identify communities for a mixed set of microgrids, some of which request external power supply while the others have excessive electricity, such that the microgrids within each community can supply their demand load by themselves regularly or when power outage occurs on the main grid.

\begin{figure*}[!htb]
	\centering \subfigure[HEC at time $t$]{\centering\includegraphics[angle=0,
		width=0.3\linewidth]{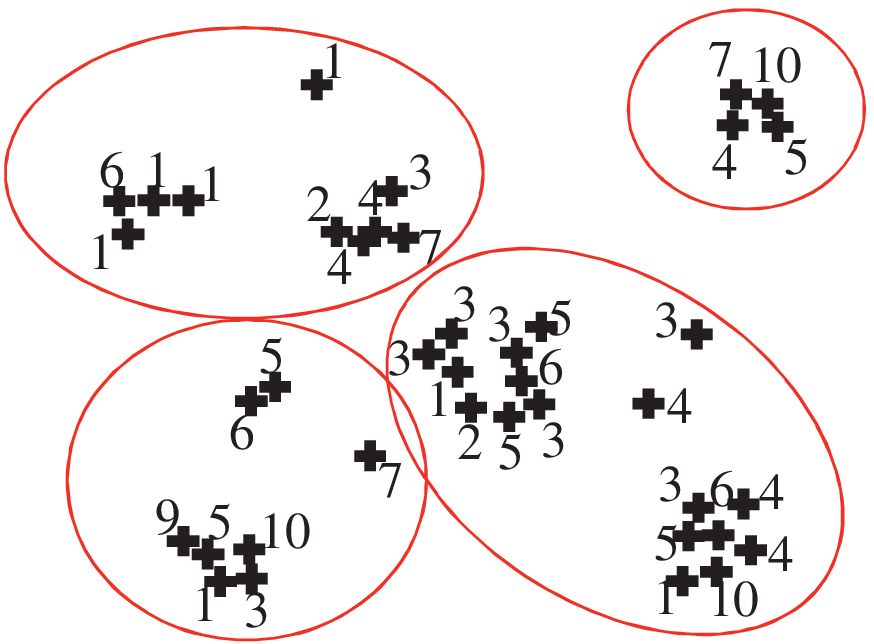} \label{fig:comm1}
	}\hspace{0.2in}\subfigure[MEC at time $t$]{\centering\includegraphics[angle=0,
	width=0.3\linewidth]{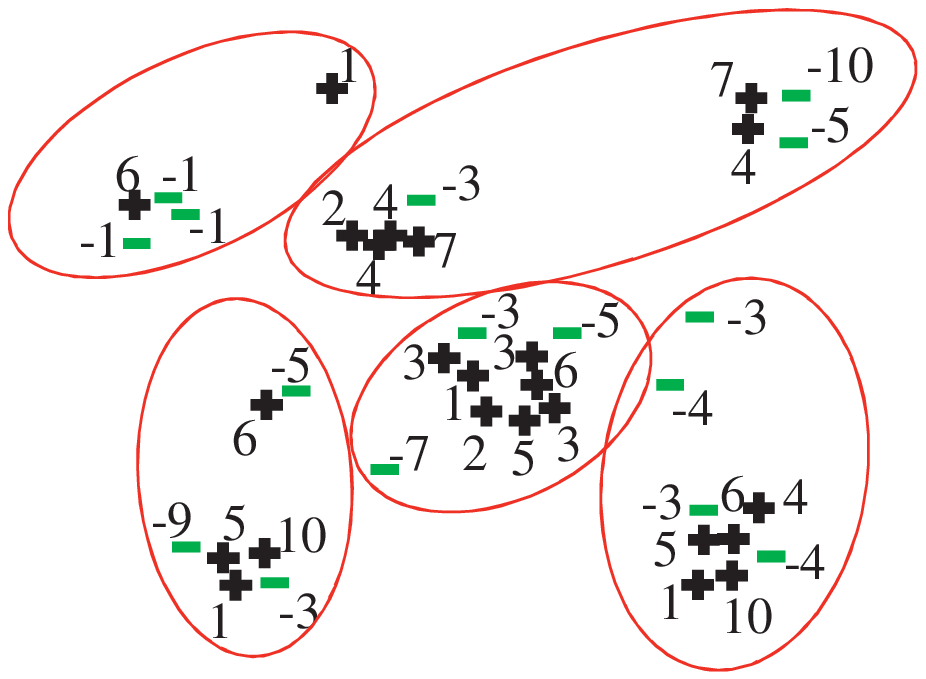} \label{fig:comm2} }
\caption[Microgrid Communities] {Energy Communities of Microgrids on the Power Grid}
\label{fig:comm}
\end{figure*}

More specifically, based on every microgrid's local energy amount (supply) and its local consumption amount (demand load), we can simply derive its \textit{Net Energy} as the amount of supply minus the demand load, which can be either \textit{positive} or \textit{negative} at specific times \footnotemark[2]. Clearly, a microgrid with positive net energy at time $t$ means that it has excessive electricity at time $t$; otherwise, it requests external power supply at time $t$. In addition, we denote the time series net energy of a microgrid $m_i$ over a period $[T_1, T_2]$ (where $T_1<T_2$) as $\forall t\in[T_1,T_2], e_i(t)$, which can be either positive or negative. 
Then, some energy communities of microgrids w.r.t. time interval $[T_1, T_2]$ can be defined as follows. 

\footnotetext[2]{If the net energy of a microgrid is 0 in $[T_1,T_2]$, then we can simply skip it or assign it to the nearest community. Thus, in this paper, we only consider the microgrids which have either positive or negative net energy.}

\subsection{Energy Communities \cite{sgc17}}

\begin{definition}[Homogeneous Energy Community (HEC)]
A group of microgrids whose net energy are exclusively positive; or exclusively negative at any time in $[T_1, T_2]$.
\end{definition}

In this case, all the microgrids in the community can feed themselves using their local energy, or all the microgrids in the community request external supply. On the contrary, if the microgrids in the community have different net energy status (positive and negative) at any time over the period $[T_1, T_2]$, we define such community as:

\begin{definition} [Mixed Energy Community (MEC)]
A group of microgrids whose aggregated net energy are mixed with positive and negative at any time in $[T_1, T_2]$. 
\end{definition}

Hence, we can categorize the energy community discovery problems based on their inputs (\emph{the net energy of all the microgrids is homogeneous or mixed between time $T_1$ and $T_2$}): (1) HECs discovery (all positive or all negative), and (2) MECs discovery (mixed with positive or negative). Figure \ref{fig:comm} presents the examples for two different energy communities on the grid at a specific time, respectively. Note that if $T_1=T_2$, HECs and MECs are obtained for a specific time instead of a time interval.

Furthermore, we define a special form of HEC or MEC in which all the microgrids' local energy can fully supply the overall demand of the community as:

\begin{definition} [Self-sufficient Energy Community (SEC)]
	A group of microgrids whose total net energy is nonnegative at any time in $[t_1, t_2]$. 
\end{definition}

In summary, we study three categories of energy community discovery problems: \emph{HECs discovery, MECs discovery} and \emph{SECs discovery}. Note that, in HECs discovery, if all the microgrids have positive net energy, then all the output HECs are automatically SECs; otherwise, all of them are not SECs. Then, we do not need to specifically identify SECs in this case. In this paper, we focus on the SECs which include the microgrids with mixed net energy (positive and negative).

\subsection{Summary of Contributions}

The energy community discovery problems are significantly different from the prior community discovery problems studied in other contexts, such as geo-locations in the spatial data \cite{communitydiscovery} and social graphs \cite{YangML13}. The key difference is that the criteria of \textit{grouping two microgrids into the same energy community should consider not only the spatial distances on the power grid but also their \emph{time series} net energy amounts (this applies to all of HECs, MECs and SECs)}. Moreover, additional constraints may apply in the problems, for example, (1) HECs may require each community to limit the overall demand load, (2) MECs and SECs may require all the microgrids in each community to balance their demand and supply, and to bound the overall net energy within a small number or even as 0 \cite{loadmanagement}, and (3) SECs require a nonnegative overall net energy for each community. In addition, both energy consumption and generation of microgrids (e.g., wind and solar) are generally stochastic, thus the energy communities (e.g., HECs, MECs, and SECs) may vary over time. How to address such issues in the energy community discovery problems? To the best of our knowledge, these have not been investigated and tackled in literature.~To address these issues, this paper has the following primary contributions:

\begin{enumerate}
	\itemsep 0.2em

	\item We define four variants of energy community discovery problems that could facilitate energy management on the power grid: \emph{two HECs discovery problems, an MECs discovery problem, and an SECs discovery problem}.

	\item We propose algorithms for the four energy community discovery problems to effectively and efficiently generate HECs, MECs, or SECs.

	\item We discuss how to apply the discovered HECs, MECs and SECs in the current energy management system, and define some utility metrics to evaluate their performance.

	\item We conduct comprehensive experiments to validate the performance of our approaches using both synthetic and real world microgrid datasets.
	
\end{enumerate} 

The rest of this paper is organized as follows. Section \ref{sec:demand}, \ref{sec:dsupply} and \ref{sec:sec} illustrate how to discover different HECs, MECs and SECs, respectively. Section \ref{sec:app} discusses how to apply the discovered communities in the current energy management system, and define some utility metrics to evaluate their performance. Section \ref{sec:exp} presents the experimental results. Section \ref{sec:discussion} discusses some issues for the communities of microgrids in the practical power grid infrastructure. 
Section \ref{sec:related} reviews the relevant literature. Finally, Section \ref{sec:con} concludes the paper and discusses the future work. 

\section{Homogeneous Energy Communities}
\label{sec:demand}
In this section, we study two different HECs discovery problems. 

\subsection{Notations}

Given $N$ microgrids $\forall i \in [1,N], m_i$, we denote its net energy at time $t$ as $e_i(t)$. If $e_i(t)<0$, then $|e_i(t)|$ represents the amount of energy demand from external resources; If $e_i(t)>0$, then $m_i$'s excessive energy amount at time $t$ is $e_i(t)$. In HECs discovery, either $\forall i\in[1,N], \forall t\in[T_1,T_2], e_i(t)<0$ holds (all negative) or $\forall i\in[1,N], \forall t\in[T_1,T_2], e_i(t)>0$ holds (all positive). For example in Figure \ref{fig:comm1}, $\forall i \in [1,N], e_i>0$, four HECs with positive net energy are circled.

Furthermore, given two microgrids $m_i$ and $m_j$, their spatial distance on the power grid is denoted as $Dis(m_i,m_j)$, which can be computed with different measures \cite{TanSK2005}, such as Euclidean distances $||m_i-m_j||_2$ and Manhattan distances $||m_i-m_j||_1$ (depending on the topology of the power grid \cite{sgsg}).

\subsection{Discovering HECs (the number of HECs is fixed)}

In real world, the smart grid may plan to partition a set of homogeneous microgrids (exclusively positive or negative) into a fixed number of HECs. For instance, the grid intends to place $K$ groups of new generators at $K$ different substations respectively to provide power supply to some newly established consumers (e.g., new constructions), then it is desirable to partition a set of microgrids with \textit{negative net energy} into $K$ different HECs such that the grid can increase the power supply to those HECs; or the grid plans to partition a set of microgrids with \textit{positive net energy} into $K$ different HECs such that the grid can establish $K$ different energy banks to store the excessive energy at different locations. 

At this time, the grid will try to identify $K$ different HECs on the power grid. Then, this energy community discovery problem can be considered as a clustering problem based on the distances of microgrids' geo-locations on the electricity transmission network, where the number of clusters is given as $K$ and the net energy of each HEC can be aggregated at different times, respectively. 

The classic K-Means algorithm \cite{macqueen1967} can efficiently generate $K$ HECs among $N$ microgrids based on \textit{their distances} on the power grid. First, $K$ locations can be arbitrarily selected as the initial centroids of $K$ HECs, then each microgrid is assigned to its nearest HEC (the nearest centroid), and finally $K$ new centroids are recomputed for the $K$ HECs. Then, repeat the above steps until convergence to generate the HECs.

For all $j\in [1,K]$, any HEC $c_j$'s net energy at time $t$ can be aggregated as $E_j=\sum_{\forall m_i \in c_j} e_i(t)$ where $t\in[T_1,T_2]$. Then, such identified HECs could help the power grid better manage their energy. For instance, if $\forall i\in[1,N], \forall t\in[T_1,T_2], e_i(t)<0$, $K$ new energy resources with power supply amount $\forall j\in [1,K], \max\{\forall t\in[T_1,T_2], |E_j(t)|\}$ will be placed at the nearest substation to the centroid of the corresponding HEC $c_j$. Then, the overall demand of each HEC (e.g., $c_j$) would not exceed $\max\{\forall t\in[T_1,T_2], |E_j(t)|\}$ at any time in $[T_1,T_2]$.

\subsection{Discovering HECs with Bounded Net Energy}

Some real world constraints may require that each HEC's net energy (either positive or negative) should be bounded, e.g., the external supply to every HEC (the negative net energy case) is limited due to capacity of generators. In these cases, the number of communities is unknown and some additional constraints should apply -- i.e. in each HEC, if the HEC's net energy is positive, then it cannot exceed a positive upper bound $L$; otherwise (negative net energy), it cannot be less than $-L$ (viz. each HEC's external demand is no greater than $L$).  

Without loss of generality, we consider the negative energy case -- the external demand of each HEC should be bounded by $L$.~To find such HECs, we extend the Density-based Spatial Clustering of Applications with Noise (DBSCAN) algorithm \cite{EsterKSX96} to ``$L^t$-DBSCAN'' by adding an upper bound $L$ for the external demand of each HEC at any time $t\in [T_1,T_2]$.~More specifically, three different microgrids can be defined \cite{EsterKSX96}:

\begin{itemize}
	
	\itemsep 0.2em 
	
	\item \textbf{Core microgrid}: a microgrid $m_i$ has at least $min$ microgrids within distance $\epsilon$ of it on the grid.
	
	\item \textbf{Reachable microgrid}: a microgrid $m_j$ is reachable from microgrid $m_i$ if there is a path $m_i,\dots, m_j$, where the next microgrid is directly reachable from the previous microgrid on the path and all the microgrids except $m_j$ are core microgrids. 
	
	Note: any two microgrids are ``neighbors'' if their distance is within $\epsilon$. Two microgrids are ``directly reachable'' means they are neighbors. For instance, ``$m_3$ is directly reachable from $m_2$, and $m_2$ is directly reachable from $m_1$'' means ``$m_3$ and $m_2$ are neighbors, and $m_2$ and $m_1$ are neighbors'', and ``$m_3$ is reachable from $m_1$''.
	
	\item \textbf{Outlier}: not reachable from any other microgrids.
\end{itemize} 

The basic idea of the DBSCAN algorithm is to group together \textit{reachable microgrids} by reaching them from the \textit{core microgrids}: scanning neighbor microgrids from the core microgrids.

However, different from the DBSCAN algorithm \cite{EsterKSX96}, discovering HECs should take into account each microgrid's external demand in the period $[T_1,T_2]$ ($\forall i\in [1,N], \forall t\in[T_1,T_2], |e_i|(t)$) as well as the spatial distances on the grid into the clustering. Our $L^t$-DBSCAN algorithm first groups microgrids based on the spatial distances between their geo-locations (similar to DBSCAN). Then, with the bounded external demand $L$ of each HEC, the $L$-DBSCAN algorithm will stop scanning microgrids for the current HEC once its aggregated external demand (at any time in $t\in [T_1,T_2]$) gets close to $L$, and then initialize a new HEC to continue scanning the microgrids based on their geo-locations. Finally, all the outliers should be assigned to their nearest communities if the updated external load (at any time in $[T_1,T_2]$) remains no greater than $L$. If no such communities found, $L^t$-DBSCAN groups the outliers to form new HECs.

Algorithm \ref{algo:cdbscan} presents the details of the $L^t$-DBSCAN algorithm for finding the HECs with negative net energy (four parameters $\epsilon, min$, $L$, and $t$). Note that $L$ can be used for bounding both positive and negative net energy.~In the algorithm, the aggregated net energy of all the microgrids in every HEC is bounded by $L$ at any time in the period $[T_1,T_2]$.

\begin{algorithm}[!h]
	\begin{algorithmic}[1]
		\small 
		\renewcommand{\algorithmicrequire}{\textbf{Input:}}
		\renewcommand{\algorithmicensure}{\textbf{Output:}}
		
		\REQUIRE $\epsilon$: distance threshold\\ 
		~~~~~$min$: core microgrid's minimum \# of ``neighbors''\\ 
		~~~~~$L$: net energy bound (e.g., external demand)\\
		~~~~~$t\in[T_1,T_2]$: time
		
		\ENSURE HECs (each HEC's external demand $\leq L$)
		
		\FOR{each microgrid $m_i$ where $i\in [1,N]$}

		\STATE get an unvisited microgrid $m_i$, and its set of neighbors (distance $\leq \epsilon$): $Close$

		\STATE initialize a new HEC with $m_i$: $c_j=\{m_i\}$ 
		
		\FOR{each microgrid $m_k$ in $Close$}

		\STATE update $c_j$'s aggregated net energy at different times: $\forall t\in[T_1,T_2], E_j(t)$

		\IF{$m_k$ is not visited}

        \STATE get $m_k$'s set of neighbors (distance $\leq \epsilon$): $Close'$

        \STATE $m_k$'s external demand at time $t$ is $|e_k(t)|$ where $t\in[T_1,T_2]$

        \IF{$\max\{\forall t\in [T_1,T_2], |E_j(t)|+|e_k(t)|\}\leq L$}

         \STATE $c=c\cup m_k$ (add $m_k$ to the HEC $c_j$)   
        
         \ELSE
         
         \STATE HEC $c_j$ is full, then go to Line 1
         
         \ENDIF

        \IF{sizeof ($Close'$)$\geq min$}
                
        \STATE $m_k$ is a core microgrid
        
        \STATE merge $Close$ and $Close'$
        
        \ENDIF
        
        \ELSE \STATE $m_k$ is a reachable microgrid
       
		\ENDIF 
		
		\ENDFOR

		\ENDFOR

	\STATE assign outliers to their nearest HECs where the aggregated external demand at any time in $[T_1,T_2]$ remains bounded by $L$. If no such communities found, group the outliers to form new HECs.

	\end{algorithmic}
	\caption{$L^t$-DBSCAN}\label{algo:cdbscan}
\end{algorithm}

\section{Mixed Energy Communities}
\label{sec:dsupply}
Among thousands of microgrids on the power grid, some of them may have excessive energy while some others may request energy from external resources (e.g., main grid). Therefore, adjacent microgrids can share their locally generated electricity for reduced energy loss on transmission and better reliability and resilience of power supply \cite{SaadHPB12,HongIJER15}. Such microgrids can form an energy community to feed their local energy demands, which are beneficial to both the power grid and themselves. Clearly, the net energy of the microgrids in the communities is mixed with negative and positive, thus called as ``Mixed Energy Communities'' (MECs). Notice that, if any two microgrids have opposite net energy status (one positive, the other one negative) at a specific time in $[T_1,T_2]$, the community discovery will be considered as MECs discovery.

The ideal case of the discovered MECs is that all the microgrids in the same MEC are geographically close to each other while balancing the local demand and supply of each MEC within a tight margin \cite{loadmanagement} (e.g., zero net energy \cite{ArboleyaGCFMSBP15,ZeroEnergy}). In Section \ref{sec:distance}, we propose an algorithm to identify such MECs on the grid towards this goal. 

\subsection{Energy-based Distance Metric}
\label{sec:distance}

Similar to the HECs, each microgrid $m_i$'s net energy at time $t$ is denoted as $e_i(t)$, which can be either \emph{positive} or \emph{negative}. While grouping two microgrids (e.g., $m_i$ and $m_j$) into an MEC, besides the spatial distance between them on the grid $Dis(m_i,m_j)$, we also have to consider their net energy $e_i$ and $e_j$ towards the load balancing of their community -- the overall demand and supply at different times should be balanced (ideally, equal to each other). For example, if one microgrid has a net energy $e_i$ while the other microgrid has a net energy demand $-e_i$, such two microgrids can supply their demands using their local energy. Thus, we define a novel measure namely ``Net Energy (NE)'' distance of two microgrids $m_i$ and $m_j$ w.r.t. time interval $[T_1,T_2]$ as: 

\begin{equation}
NE(m_i,m_j)= \sum_{t=T_1}^{T_2}|e_i(t)+e_j(t)|
\end{equation}

If $\forall t\in[T_1,T_2]$, $e_i(t)+e_j(t)=0$ hold, then we have $NE(m_i,m_j)=0$. However, if $\forall t\in[T_1,T_2]$, $e_i(t)=e_j(t)$ hold, we have $NE(m_i,m_j)$=$2\sum_{t=T_1}^{T_2}|e_i|$. The NE distance differs from other distance measures used in traditional community discovery problems due to its unique feature: two opposite values (e.g., $e_i$ and $-e_i$) are measured as ``close''.

\subsection{Algorithm}

For the MECs discovery, we define \emph{two maximum distance thresholds} for the normalized NE distances and the normalized spatial distances respectively: $\epsilon,\epsilon'\in [0,1]$. Then, we propose a novel agglomerative algorithm \cite{TanSK2005} to identify MECs by utilizing $\epsilon$ and $\epsilon'$ to specify the criteria for bounding the differences between the overall supply and demand of each community and the spatial distances between the microgrids in each community. Specifically, we let each microgrid find its nearest microgrid (with an NE distance $\leq\epsilon$ and a spatial distance $\leq\epsilon'$) to form an MEC, update the MEC centroid's geo-location and net energy, and then hierarchically merge ``small MECs'' to form ``large MECs'' (for \emph{pursuing better resilience}). The merging process terminates if the NE distance between any two MECs' centroids exceeds $\epsilon$ or their spatial distance exceeds $\epsilon'$. Algorithm \ref{algo:agg} presents the details.

\begin{algorithm}[!h]
	\begin{algorithmic}[1]
		
		\small 
		\renewcommand{\algorithmicrequire}{\textbf{Input:}}
		\renewcommand{\algorithmicensure}{\textbf{Output:}}
		\REQUIRE $\epsilon$: maximum threshold of the NE distances\\
		~~~~~$\epsilon'$: maximum threshold of the spatial distances
		
		\ENSURE MECs
		
		\WHILE{any ungrouped microgrid $m_i$ in $m_1,\dots, m_N$} 
		
		\STATE initialize a new MEC with $m_i$: $c_j = \{m_i\}$ 

		\FOR{each ungrouped microgrid $m_k$}

	    \STATE compute MEC $c_j$'s net energy at time $\forall t\in[T_1,T_2]$: $E_j(t)$ and its centroid's geo-location $\mu_j$

		\IF{$NE(\mu_j,m_k)\leq\epsilon$ and $Dis(\mu_j,m_k)\leq\epsilon'$}

		\STATE $c_j = c_j \cup m_k$ (add $m_k$ to the MEC $c_j$)

		\STATE update $\forall t\in[T_1,T_2], E_j(t)$ and $\mu_j$

		\ENDIF \ENDFOR \ENDWHILE
		
		\STATE considering each MEC $c_j$ as a microgrid with net energy $E_j(t)$ at time $t$ and geo-location $\mu_j$, repeat Line 1-7 to hierarchically merge the MECs based on $\epsilon$ and $\epsilon'$ until convergence
		
	\end{algorithmic}
	\caption{Two-threshold MECs Discovery}\label{algo:agg}
\end{algorithm}

Therefore, the difference of the overall supply and demand of every MEC is bounded/balanced at different times by $\epsilon$, and the spatial distance between any microgrid and its MEC's centroid is bounded by $\epsilon'$.

\section{Self-sufficient Energy Communities}
\label{sec:sec}
Many real world applications require that the microgrids in each MEC can fully supply their demand with their local energy (e.g., large-scale blackouts). Therefore, it is also desirable to discover the ``Self-sufficient Energy Communities (SECs)'' with \textit{nonnegative} net energy \cite{CommunityEnergy}. In this section, we present two approaches for discovering SECs, which are special MECs on the grid.

Specifically, given $N$ microgrids $m_1,\dots, m_N$, we denote the number of SECs for the $N$ microgrids as $K$. Then, the $K$ SECs can be denoted as $c_1,\dots, c_K$, and we can define binary variables $\forall i\in [1,N], \forall j\in [1,K], x_{ij}\in\{0,1\}$ to indicate if microgrid $\forall i\in[1,N], m_i$ is included in SEC $\forall j\in [1,K], c_j$ or not: if $x_{ij}=1$, microgrid $m_i\in c_j$; otherwise, not. We now mathematically formulate the problem of discovering SECs.  

\subsection{Optimization-based SECs Discovery}

If the aggregated net energy of the given microgrids is non-negative in $[T_1,T_2]$, we can formulate an optimization problem for discovering SECs. 

\subsubsection{Clustering Constraints}

First, every microgrid can only be assigned to exactly one SEC. This criterion creates a group of clustering constraints as below:

\begin{equation}
\forall i\in [1,N], \sum_{\forall j=1}^K x_{ij}=1
\end{equation}

Second, recall that the net energy of any SEC should be non-negative at any time $t\in [T_1,T_2]$. This criterion creates another group of clustering constraints as below:

\begin{equation}
\forall t\in[T_1,T_2], \forall j\in[1,K], \sum_{\forall i=1}^N[e_i(t)x_{ij}]\geq 0\\
\end{equation}

Then, the clustering constraints of SECs can be summarized as below:

\begin{equation}
\begin{gathered}
s.t.{\centering
	\begin{cases}
	\forall i\in [1,N], \sum_{\forall j=1}^K x_{ij}=1\\
	\forall t\in[T_1,T_2], \forall j\in[1,K], \sum_{\forall i=1}^N[e_i(t)x_{ij}]\geq 0\\
	\forall i\in[1,N], \forall j\in[1,K], x_{ij}\in\{0,1\}
	\end{cases}}
\end{gathered}\label{eq:cont}
\end{equation}

\subsubsection{Problem Formulation}
\label{sec:optp}

If all the binary variables $\forall x_{ij}$ satisfy all the constraints in Equation \ref{eq:cont}, all the output energy communities would be SECs. Thus, we can solve the constraint satisfaction problem (CSP) without an objective function to find out feasible solutions w.r.t. SECs. Note that such constraint satisfaction problem is NP-hard due to the involvement of a large number of binary variables. 

More importantly, besides the constraint satisfaction problem, we can formulate the SECs discovery problem by minimizing the overall load on the transmission lines (energy loss in transmission) in all the SECs. Then, we can denote the energy loss rate as $\theta$: e.g., transmitting an amount of energy $100$ (Watts), the load on 1 unit distance is $100\theta$ (Watts). W.o.l.g., given microgrid $m_i$ with positive net energy at time $t$: $e_i(t)$ and any other microgrid $m_s$ with negative net energy at time $t$: $e_s(t)$, we define the amount of energy from $m_i$ to $m_s$ at time $t$ as $y_{is}(t)$. Thus, the overall load on the transmission lines can be represented using the model in \cite{HongIJER15}: 

\begin{equation}
\sum_{t=T_1}^{T_2}\sum_{j=1}^{K}\sum_{i=1}^N\sum_{s=1,s\ne i}^N [x_{ij}x_{sj}y_{is}(t) \theta Dis(m_i,m_s)]
\label{eq:load}
\end{equation}

If $x_{ij}=1$ and $x_{sj}=1$ (viz. $m_i,m_s\in c_j$), then the load of the power flow from $m_i$ to $m_s$ at time $t$ is $y_{is}(t) \theta Dis(m_i,m_s)$. If $x_{ij}$ or $x_{sj}=0$ (viz. they are not in the same community), there is no power transmission from $m_i$ to $m_s$, and the load is $0$. Then, the overall load on the transmission lines can be aggregated as Equation \ref{eq:load}. In the meanwhile, there are two additional sets of power flow constraints: 

\begin{equation}
\begin{gathered}
s.t.{\centering
	\begin{cases}
	\forall t, \forall i \in[1,N], \sum_{s=1,s\ne i}{N}[x_{ij}x_{sj}y_{is}(t)] \leq e_i(t)\\
	\forall t, \forall s\in[1,N], \sum_{i=1, i\ne s}^N[x_{ij}x_{sj}y_{is}(t)](1-\theta)\geq |e_s(t)|\\
	\forall t, \forall i, \forall s\in[1,N], y_{is}(t)\geq 0
	\end{cases}}
\end{gathered}\label{eq:PPEE}
\end{equation} 

where the above two sets of constraints ensures that the overall outgoing energy of every microgrid with positive net energy is no greater than its current excessive energy, and the overall incoming energy of every microgrid with negative energy is no less than its current demand, respectively \cite{HongIJER15}. In summary, we consider Equation \ref{eq:load} as the objective function, and combine Equation \ref{eq:cont} and \ref{eq:PPEE} as constraints

\begin{equation}
\begin{gathered}
\min: \sum_{t=T_1}^{T_2}\sum_{j=1}^{K}\sum_{i=1}^N\sum_{s=1,s\ne i}^N [x_{ij}x_{sj}y_{is}(t) \theta Dis(m_i,m_s)]\\
s.t.{\centering
	\begin{cases}
	\forall i\in [1,N], \sum_{\forall j=1}^K x_{ij}=1\\
	\forall t, \forall j\in[1,K], \sum_{\forall i=1}^N[e_i(t)x_{ij}]\geq 0\\
	\forall t, \forall i \in[1,N], \sum_{s=1,s\ne i}{N}[x_{ij}x_{sj}y_{is}(t)] \leq e_i(t)\\
	\forall t, \forall s\in[1,N], \sum_{i=1, i\ne s}^N[x_{ij}x_{sj}y_{is}(t)](1-\theta)\geq |e_s(t)|\\
	\forall i\in[1,N], \forall j\in[1,K], x_{ij}\in\{0,1\}\\
	\forall t\in [T_1,T_2], \forall i, \forall s\in[1,N], y_{is}(t)\geq 0
	\end{cases}}
\end{gathered}\label{eq:opt2}
\end{equation}

\subsubsection{Tabu Search based Algorithm}

Due to the NP-hardness of the optimization problem, we propose a Tabu Search \cite{tabu1990} based meta-heuristic algorithm to solve the problem. Specifically, the algorithm first specifies a range for the number of SECs $K\in\{K_{min},\dots, K_{max}\}$, and arbitrarily partitions all the microgrids into $K$ groups based on their geo-locations. Then, for every $K\in\{K_{min},\dots, K_{max}\}$, the algorithm iteratively searches the neighboring solutions to make the number of SECs reach $K$ where ``moving a microgrid from one group to another nearest group'' is defined as one of its neighboring solutions. After obtaining a set of candidate neighboring solutions (different moves), the neighboring solution which can improve the objective function most (reduce the load with the greatest amount), then replace the current solution with the neighboring solution. To improve the performance of searching performance, the following criteria are integrated in the algorithm:

\begin{itemize}
	\item To avoid the solutions getting stuck in local optimum while searching SECs for every $K$, a Tabu list is defined with length $S$ which stores $S$ most recent solutions that replaced th previous solution. Then, in the searching process, if any neighboring solution is found in the Tabu list, the searching process continues without visiting such neighboring solution. 
	
	\item Among all the SECs, select the SEC with the highest net energy (positive) at most times in $[T_1,T_2]$, and then move each microgrid with the positive net energy to the corresponding nearest non-SEC, such that a set of candidate neighboring solutions can be found.
	
\end{itemize}

The load based objective function cannot be reduced for the current $K$. Then, the algorithm moves to the next $K\in\{K_{min},\dots, K_{max}\}$. Among all the discovered SECs for all $K\in \{K_{min},\dots, K_{max}\}$, the best solution (with the minimum overall load on the transmission lines while satisfying all the constraints) will be selected as the output SECs.

\subsection{A Two-phase Algorithm for Discovering SECs}

Besides the optimization-based approach -- which formulates the optimization problem and solves the problem with a Tabu Search based algorithm, we present a two-phase algorithm to discover a subset of microgrids to form the SECs. Notice that, if the overall net energy of all the given microgrids are \emph{negative} in $[T_1,T_2]$, the constraints in the optimization based approach cannot be satisfied simultaneously to form the SECs for all the given microgrids. Instead, the proposed two-phase heuristic algorithm can still effectively discover SECs out of the given microgrids.

Specifically, among all the $N$ microgrids, we denote the set of microgrids with \emph{positive net energy at any time in $[T_1,T_2]$} as $M^+$, and the set of microgrids with \emph{any negative net energy in $[T_1,T_2]$} as $M^-$. Then, the two phases are illustrated as follows.

\begin{itemize}
	\item Phase (1): the algorithm first clusters all the microgrids in $M^+$ based on their geo-locations, where each cluster can be considered as a ``merged microgrid'' with aggregated positive net energy. In this stage, we extend the $K$-Means algorithm \cite{macqueen1967} to cluster such microgrids' geo-locations by specifying different $K\in\{K_{min},\dots, K_{max}\}$. Then, the algorithm repeats $K$-Means with different $K$ values and chooses the best clustering result -- the minimum sum of squared errors (SSE) of the spatial distances \cite{TanSK2005} in all the clustering results.
	
	\item Phase (2): denoting the clustering result of $M^+$ as $c_1^*,\dots, c_K^*$, the net energy of any cluster $\forall j\in[1,K]$, $c_j^*$ at time $t$ can be aggregated as $\sum_{\forall m_i\in c_j^*}e_i(t)$. Then, $\forall j\in[1,K], c_j^*$ iteratively adds its centroid's nearest ungrouped microgrid in $M^-$ until its net energy drops close to $0$ at any time in $[T_1,T_2]$ 
\end{itemize}

Finally, the updated $c_1^*,\dots, c_K^*$ are identified as $K$ different SECs. The details of the two-phase algorithm are given in Algorithm \ref{algo:sscd}. Note that Algorithm \ref{algo:sscd} involves all the microgrids in $M^+$ in the SECs, but may not involve all the microgrids in $M^-$  (depending on the net energy of the microgrids in $M^+$ and $M^-$). Furthermore, the net energy of most self-sufficient communities can be well balanced to form ``Zero Net Energy'' communities \cite{ZeroEnergy}.

\begin{algorithm}[!h]
	\begin{algorithmic}[1]
		\small 
		\renewcommand{\algorithmicrequire}{\textbf{Input:}}
		\renewcommand{\algorithmicensure}{\textbf{Output:}}
		\REQUIRE $M^+$: set of microgrids with positive net energy\\ 
		~~~~~$M^-$: set of microgrids with negative net energy\\ 
		~~~~~$\{K_{min},\dots, K_{max}\}:$ possible values for $K$
		
		\ENSURE SECs
		
		\FOR{$K=K_{min},\dots,K_{max}$}
		
		\STATE run $K$-Means for all microgrids in $M^+$ based on their geo-locations to obtain $c_1,\dots, c_K$
		
		\ENDFOR
		
		\STATE choose the best clustering result with the minimum SSE for different $K$: $c_1^*,\dots, c_K^*$ (best $K$)
		
		\FOR{$j\in[1,K]$}
		
		\STATE compute the centroid of $c_j^*$ as $\mu_j^*$
		
		\WHILE{$\forall t\in[T_1,T_2], \sum_{\forall m_i\in c_j^*}e_i(t)\geq 0$}
		
		\STATE find $\mu_j^*$'s nearest ungrouped microgrid in $M^-$, denoted as $m_k$
		
		\STATE $c_j^*=c_j^*\cup m_k$ (add $m_k$ to the SEC $c_j$)
		
		\STATE update $c_j^*$'s net energy: $\forall t\in[T_1,T_2], \sum_{\forall m_i\in c_j^*}e_i(t)+=e_k(t)$ and the geo-location of $\mu_j^*$
		
		\ENDWHILE
		
		\ENDFOR
		
     \STATE return the updated $c_1^*,\dots, c_K^*$ as SECs
	\end{algorithmic}
	
	\caption{Two-phase SECs Discovery}\label{algo:sscd}
\end{algorithm}

\section{Utilities of HECs, MECs and SECs}
\label{sec:app}
In this section, we discuss how to apply the discovered energy communities on the power grid, and present some utility metrics to evaluate the output energy communities. 

\subsection{HECs}

\begin{figure}[!tbh]
	\centering
	\includegraphics[angle=0, width=0.45\linewidth]{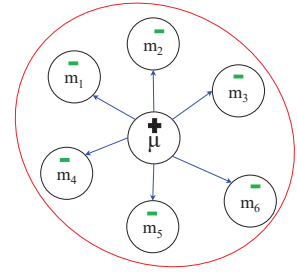} 
	\caption[Optional caption for list of figures]
	{Star Structure of Each HEC (Does Not Change Over Time)}
	\label{fig:hec}
\end{figure}

Microgrids in each of the heterogeneous energy communities can be considered as a larger supplier (the case that all the microgrids have positive net energy) or a larger consumer (the case that all the microgrids have negative net energy). All the microgrids in each HEC will be interconnected in a star structure all the time (since the net energy of all the microgrids in the same community is always homogeneous). For instance, in the negative net energy case, additional power supply (e.g., generators) will be placed at the centroid of each HEC (as shown in Figure \ref{fig:hec}). 

Then, given microgrids $\forall m_i$ in the HEC $c_j$ and $m_i$'s external demand at time $t$: $|e_i(t)|$ where $j\in[1,K]$, the energy transmission amount in the power flow at time $t$ within the community will be determined as: $\forall m_i\in c_j$, transmitting energy with the amount $|e_i(t)|$ from $\mu_j$ (centroid) to $m_i$. Furthermore, we can identify some utility metrics for evaluating HECs as below.

\begin{itemize}
	
	\itemsep 0.2em
	
	\item \emph{Average distance between each microgrid to its centroid}: shorter distance could reduce the energy loss during transmission between each microgrid and its centroid. We can use the metric of (spatial) sum of squared errors (SSE) of all the HECs\cite{WuLXCC15} to measure such average distance.
	
	\item \emph{The average net energy of each HEC by taking account into each microgrid's net energy at different times in $[T_1,T_2]$}: denoting $|t|$ as the number of timestamps in $[T_1,T_2]$ utilized for energy community discovery, identifying HECs based on the microgrids w.r.t. more timestamps (viz. larger $|t|$, longer $[T_1,T_2]$) would reflect more accurate results of HECs.
	
	\item \emph{The load on transmission lines}: HECs have better utility if such load is lower.
	
\end{itemize}

\subsection{MECs and SECs}

\begin{figure}[!tbh]
	\centering
	\includegraphics[angle=0, width=0.55\linewidth]{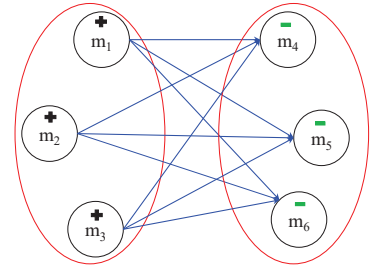} 
	\caption[Optional caption for list of figures]
	{Bipartite Graph of Each MEC or SEC (May Change Over Time)}
	\label{fig:mec}
\end{figure}

After discovering MECs and SECs, microgrids could cooperate with each other by sharing their local energy \cite{HongIJER15}. Since every microgrid can only be either a power supplier or consumer \cite{SaadHPB12}, MECs and SECs are implemented as a bipartite graph on the power grid: in each MEC or SEC, power might be routed from any microgrid with positive net energy to any microgrid with negative net energy (as shown in Figure \ref{fig:mec}). 

Note that, the structure of the bipartite graph may change over time (e.g., $M_1$ might be a supplier at time $T_1$ and it may become a consumer at time $T_2$). Also, the connection between every pair of microgrids can be available via the power transmission network of the main grid \cite{FangMXY12}. As illustrated in Section \ref{sec:optp}, the optimal energy transmission solution (power flow) within each community can be obtained using the model in \cite{HongIJER15} (which is simplified from the optimization model in Section \ref{sec:optp}):

\begin{equation}
\begin{gathered}
\min: \sum_{\forall i, \forall s} y_{is}(t)\\
s.t.{\centering
	\begin{cases}
	\forall i, \sum_{\forall s}y_{is}(t) \leq e_i(t)\\
	\forall s, \sum_{\forall i}y_{is}(t)(1-\theta)\geq |e_s(t)|\\
	\forall i, \forall s, y_{is}(t)\geq 0
	\end{cases}}
\end{gathered}\label{eq:PPEE2}
\end{equation}

Note that SECs definitely have optimal solutions in the above problem. If MECs cannot find an optimal solution (overall demand exceeds overall supply in any MEC), the main grid will fill the gap \cite{HongIJER15}. Similarly, we can also identify some utility metrics for evaluating MECs and SECs as below.

\begin{itemize}
	\itemsep 0.2em
	
	\item \emph{Average distance between every pair of power supplier (positive net energy) and consumer (negative net energy)}: shorter distance could reduce the energy loss during transmission from the power supplier to the power consumer. Since the structure of the bipartite graph may change over time, we still use the metric of the (spatial) SSE of all the communities to measure such average distance.
	
	\item \emph{Ratio of nonnegative MECs in $[T_1,T_2]$}: the MECs with higher ratio of nonnegative net energy could generate more SECs. If an MEC has low ratio of nonnegative MECs, external supply from the main grid is still required. 
	
	\item \emph{The average net energy of each MEC or SEC by taking account into each microgrid's net energy at different times in $[T_1,T_2]$}: denoting $|t|$ as the number of timestamps utilized for energy community discovery, we identify MECs and SECs based on the microgrids' energy status at more timestamps (e.g., larger $|t|$, longer period $[T_1,T_2]$) would reflect more accurate results of the communities.
	
	\item \emph{The load on transmission lines}: MECs and SECs have better utility if such load is lower.
	
\end{itemize}

\section{Experiments}
\label{sec:exp}

\subsection{Experimental Setup}

\noindent\textbf{Datasets.} Our experimental simulations were conducted on the synthetic data generated from three real world datasets: a spatial dataset and two power generation \& consumption datasets. First, the spatial dataset of 115,475 cities/towns in the U.S. was collected by the US Geological Survey on July 7, 2012 and is available in National Imagery and Mapping Agency \cite{uwater}. Second, two power generation \& consumption datasets were collected by Richardson et al. \cite{Richardson20101878} in East Midlands, UK, and Barker et al. \cite{umassdata} in Massachusetts, US. Specifically, Richardson et al. \cite{Richardson20101878} collected 22 dwellings' power consumption over 2 years. Barker et al. \cite{umassdata} collected a low resolution dataset (\textit{UMass Smart* Home Dataset}): 443 households' power consumption on April 2, 2011, and a high resolution dataset (\textit{UMass Smart* Microgrid Dataset}): three microgrids' power generation \& consumption over 3 months in 2012. In the UMass Smart* Microgrid dataset, both solar panels and wind turbines were installed in three microgrids.

In our experiments, we generated synthetic datasets based on the real world spatial dataset, and the time series generation \& consumption datasets:

\begin{enumerate}
	
	\itemsep 0.2em 
	
	\item We first aggregated all the real world generation and consumption datasets with the frequency of one reading per 15 minutes.
	
	\item To test the HECs, we generated a synthetic dataset by sampling 50,000 microgrids' power consumption over 1 month based on the 22 real dwellings' power consumption data \cite{Richardson20101878}, and then randomly assigning geo-locations in the spatial dataset \cite{uwater} to the 50,000 microgrids. 
	
	\item To test the MECs, we generated two synthetic datasets by sampling 50,000 microgrids' power generation and consumption over 1 month based on the microgrid dataset in \cite{umassdata}, and then randomly assigning geo-locations in the spatial dataset \cite{uwater} to the 50,000 microgrids. 
	
	\item To test the SECs, we used the data in (3) MECs discovery to evaluate the two-phase algorithm. To compare the optimization-based approach and the two-phase algorithm, we selected 10,000 microgrids with a high percent of microgrids with positive net energy out of the above 50,000 microgrids with both generation and consumption -- ensuring that the optimization based SECs discovery can find a feasible solution.
	
\end{enumerate}

Note that two microgrids may have the same geo-location in our experiments (such case exists in real world: two households might be neighbors). Furthermore, we converted all the real world power generation \& consumption amounts extracted from the datasets to power rates (in Watts). Then, there are $1.44\times 10^8$ power consumption rates ($5\times 10^4$ microgrids, $2,880$ timestamps per microgrid) generated in each of the synthetic datasets. Table \ref{table:data} shows the characteristics of the datasets. 

\begin{table}[!h]
 
	\centering \caption{Characteristics of Datasets} 	
	\begin{tabular}{|c|l|}
		\hline
		\small{\textbf{Datasets}} & ~~~~~~~~~~~~~~~~~\small{\textbf{Characteristics}} \\
		\hline
		
		Spatial Data & 115,475 unique geo-locations \\
		\hline

		UK Power Consumption & average consumption rate: 1,172\\
	(Watts)& max consumption rate: 2,891\\ 
		& min consumption rate: 140\\

		\hline
		
		Synthetic Data for HECs & 50,000 microgrids (all negative)\\
		& number of timestamps: 2,880\\
		\hline

		UMass Smart* Microgrid & average generation rate:  921\\
		Generation \& Consumption& average consumption rate:  1,368\\
		(Watts) & max generation rate: 1,250\\ 
		 & max consumption rate: 2,147\\ 
		& min generation rate: 355\\
		& min consumption rate: 192\\
		
		\hline
		
		Synthetic Data for MECs & 50,000 microgrids\\
		
		(also for SECs Discovery & 25,317 microgrids (all $+$ in $[T_1,T_2]$)\\
		
		using Two-phase Algorithm) & 24,683 microgrids (existing $-$ in $[T_1,T_2]$)\\
		
		  & number of timestamps: 2,880\\
		
		\hline
		
		Synthetic Data for SECs & 10,000 microgrids\\
		
	  	(comparing two algorithms)& 6,588 microgrids (all $+$ in $[T_1,T_2]$) \\
		
		&3,412 microgrids (existing $-$ in $[T_1,T_2]$)\\
		
		& number of timestamps: 2,880\\
		
		\hline

	\end{tabular}
	\label{table:data}
\end{table}

\begin{figure*}[!tbh]
	
	\subfigure[Microgrids \# in HECs vs. Number of HECs $K$]{
		\includegraphics[angle=0, width=0.45\linewidth]{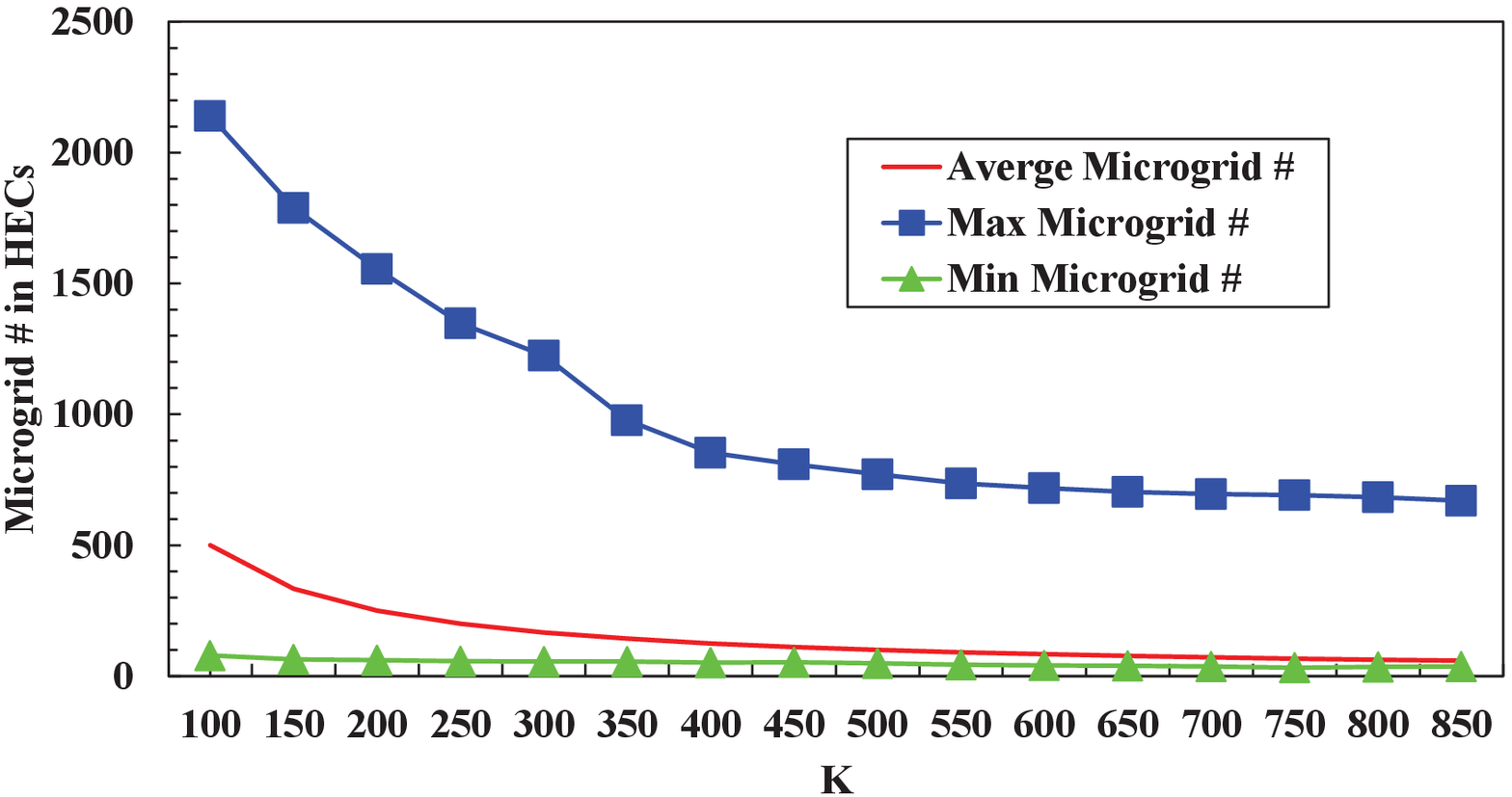}
		\label{fig:kmm} }
	\hspace{0.1in}
	\subfigure[External Demand of HECs vs. Number of HECs $K$]{
		\includegraphics[angle=0, width=0.45\linewidth]{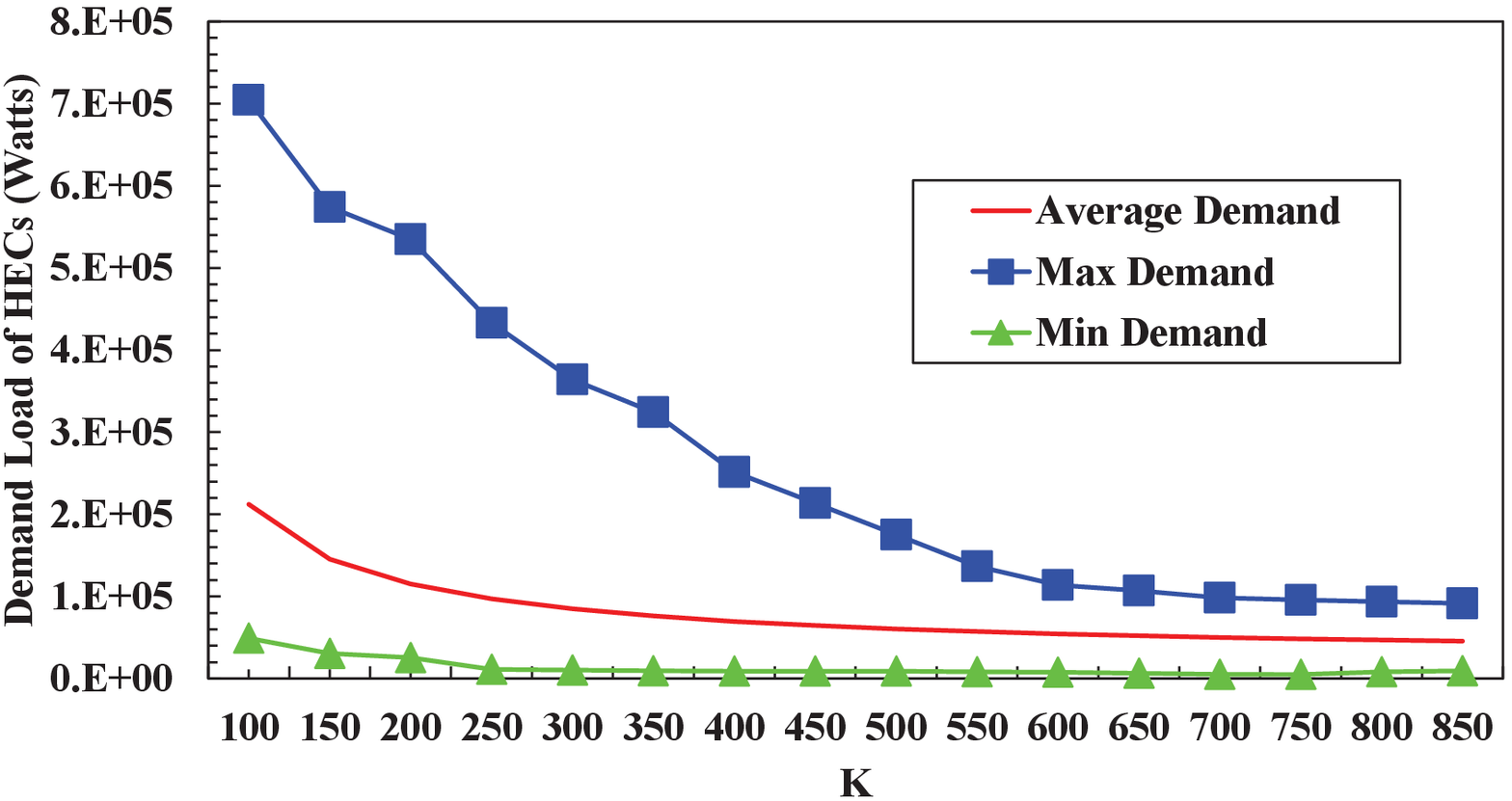}
		\label{fig:kml} }  
	\caption[Optional caption for list of figures]
	{HECs Discovery with Fixed Number of Communities $K$ (50,000 Microgrids, 2,880 Timestamps: $|t|=2880$)}
	\label{fig:km}
\end{figure*}

\begin{figure*}[!tbh]
	\subfigure[Microgrids \# in HECs vs. Net Energy Bound in Each HEC $L$]{
		\includegraphics[angle=0, width=0.45\linewidth]{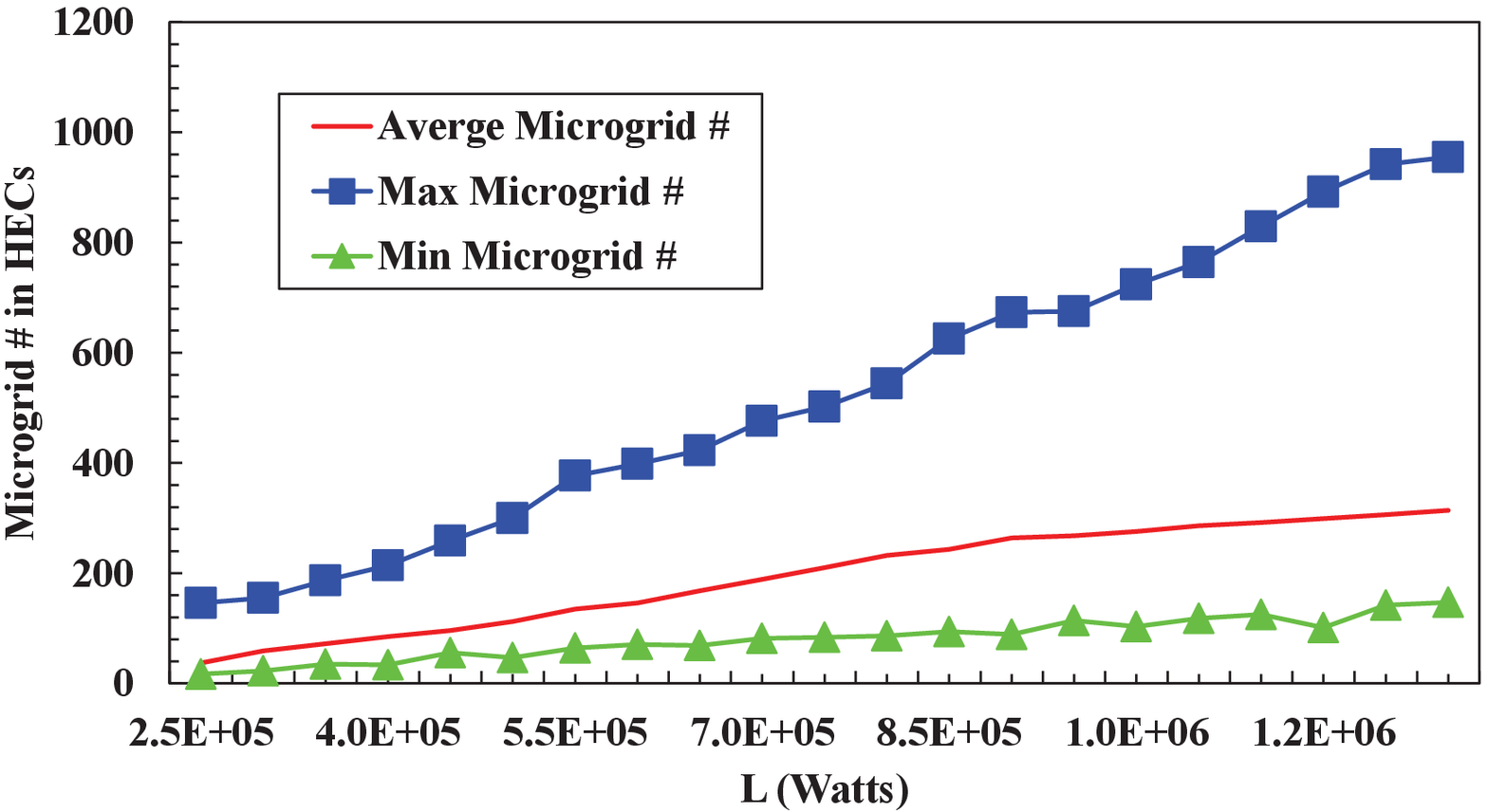}
		\label{fig:dbm} } 
	\hspace{0.1in} 
	\subfigure[External Demand of HECs vs. Net Energy Bound in Each HEC $L$]{
		\includegraphics[angle=0, width=0.45\linewidth]{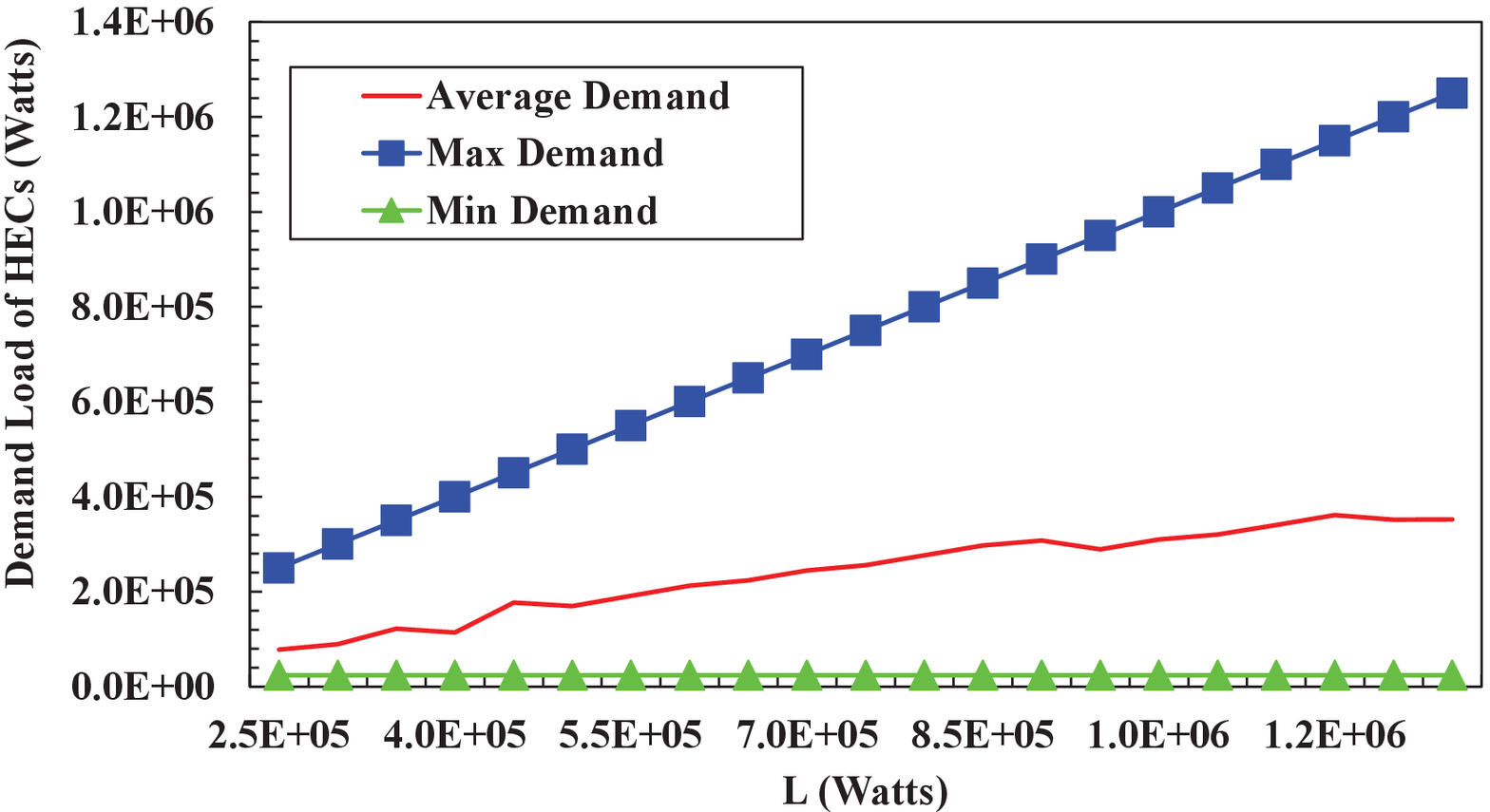}
		\label{fig:dbl} } 
	
	\caption[Optional caption for list of figures]
	{HECs Discovery with Bounded Net Energy $L$ (50,000 Microgrids, 2,880 Timestamps: $|t|=2880$)}
	\label{fig:db}
\end{figure*}

\vspace{0.05in}

\noindent{\bf Normalization.} We use Euclidean distance to measure the spatial distance between any two microgrids on the grid. Both the Euclidean distances and the net energy (NE) distances are normalized into $[0,1]$ in all the experiments.

\vspace{0.05in}

\noindent{\bf Platform.} All the experiments were simulated on a DELL PC with Intel Core i7-4790 CPU 3.60GHz and 16G RAM running Microsoft Windows 8.1 Operating System.

\subsection{Discovering HECs}

The algorithms have an identical performance to discover HECs with negative and positive net energy. W.l.o.g., we evaluate the case of negative net energy (external demand).

\subsubsection{Discovering K HECs}

We first apply the $K$-Means algorithm \cite{KanungoMNPSW02} to cluster the 50,000 microgrids based on their Euclidean distances, and then aggregate the external demand in each HEC. In literature, the performance of the $K$-Means algorithm on clustering has been well studied using measures such as (spatial) sum of squared errors (SSE) \cite{WuLXCC15} and silhouette coefficient \cite{TanSK2005} to evaluate the cohesion and separation of the clusters. Therefore, we do not report the spatial cohesion and separation of the HECs on the grid here. 

Figure \ref{fig:km} shows the sizes and the external demands of the HECs. More specifically, Figure \ref{fig:kmm} presents the average, maximum and minimum number of microgrids in all the HECs (parameter $K$ varies in $[100,850]$). With an increasing $K$, the average, maximum and minimum number of microgrids in all the HECs have a descending trend in general (simply because the number of HECs increases). 

Note that the average, maximum and minimum external demands of the HECs in Figure \ref{fig:kml} are derived from the microgrids' external demand w.r.t. all 2,880 timestamps. As a result, the average, maximum, and minimum external demands of the HECs also decline as $K$ increases. Note that if $K=100$, a large HEC ($\sim 2,200$ microgrids) can be identified to request external energy (with an amount $\sim 10^6$ Watts), then the size and external demands of the HECs drop significantly as $K$ increases. In this case, the $K$-Means algorithm is applied to discover HECs only based on the geo-locations of microgrids. A different $|t|$ (utilizing energy amounts at $|t|$ different timestamps for communities discovery) does not affect the results of discovered HECs.

Finally, we let $\theta=0.0001$ per 0.1 (normalized spatial distance), applied $K$-Means to simulate five substations of the main grid, and derive the average distance between each of the 50,000 microgrids and the main grid (its nearest substation). We then compare the overall load on transmission lines at 2,880 timestamps for 50,000 microgrids in two cases: with HECs and without HECs. Table \ref{table:load} shows that the energy loss during transmission can be significantly reduced with HECs. 

\begin{table}[!h]

	\centering \caption{Load on Transmission Lines ($K$-Means)} 
	\begin{tabular}{|c|c|c|c|c|}
		\hline
		\small{$K$}  & 100 & 300  & 500 & 700  \\
		\hline
		
		Load (with HECs) & 1576913 & 463552 & 217717 & 136330 \\
		Load (without HECs) & 4951490 & 4951490 & 4951490 & 4951490\\
		\hline

	\end{tabular}
	\label{table:load}
\end{table}

\subsubsection{Discovering HECs with Bound $L$ at Time $t\in [T_1,T_2]$}

\begin{figure}[!h]
	\centering 	
	\subfigure[Average External Demand w.r.t. Different Number of Timestamps $|t|$]{
		\includegraphics[angle=0, width=0.9\linewidth]{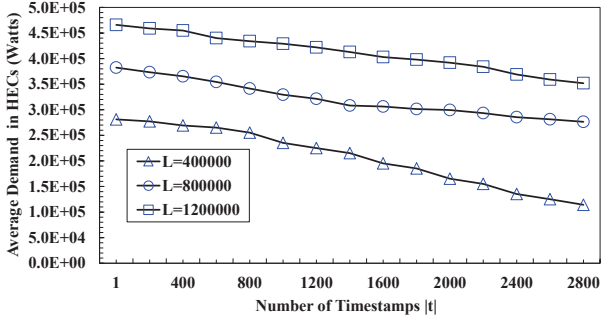}
		\label{fig:dbave} } 
	\subfigure[SSE of the HECs w.r.t. Different Number of Timestamps $|t|$]{
		\includegraphics[angle=0, width=0.9\linewidth]{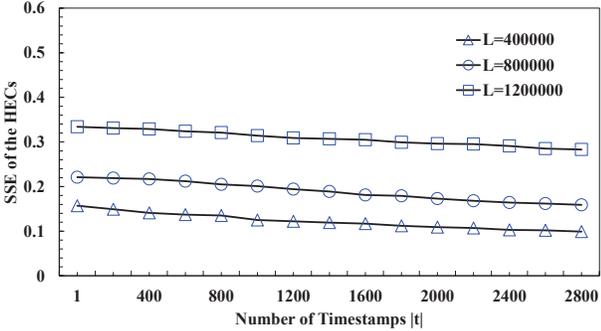}
		\label{fig:dbsse} } 
	\caption[Optional caption for list of figures]
	{HECs Discovery w.r.t. Different $|t|$ (50,000 Microgrids)}
	\label{fig:time}
\end{figure}

Using $L^t$-DBSCAN to discover HECs with negative energy, the external demand in every HEC is bounded by $L$ (Watts) at any time in $[T_1,T_2]$. Similar to the $K$-Means, we do not report the spatial cohesion and separation of the HECs generated by the $L^t$-DBSCAN algorithm with different parameters $\epsilon$ and $min$. Instead, We set a reasonable value for the normalized minimum distance (Euclidean) $\epsilon=0.1$ and the core microgrid's minimum number of neighbors $min=10$.

In the experiments, we implemented Algorithm \ref{algo:cdbscan} with a varying $L$ and all the 2,880 timestamps. Note that outliers in the algorithm were either assigned to the nearest HECs or grouped separately. Also, every HEC's external demand does not exceed $L$ at any time. Then, we plotted the average, maximum and minimum number of microgrids in all the HECs w.r.t. all the 2,880 timestamps in Figure \ref{fig:dbm}. The maximum number of microgrids in a single HEC has a near-linear increasing trend as $L$ increases. Compared to the maximum number of microgrids, the average and minimum numbers of microgrids in all the HECs increase extremely slowly. In addition, Figure \ref{fig:dbl} shows the average, maximum and minimum external demand of all the HECs w.r.t. all the 2,880 timestamps. The maximum external demand of all the HECs always equals $L$ since the net energy bound $L$ is the major constraint besides the distances of microgrids' geo-locations. However, the average and minimum external demands of all the HECs tend flat as $L$ increases. In reality, the HEC with the minimum demand only includes a small number of microgrids, where not many microgrids can be reachable from other microgrids. Thus, the minimum external demand of such HEC would be far less than $L$ at different times.

Furthermore, we have some other interesting findings in the HECs discovery by utilizing microgrids' time series external demand over different lengths of periods (varying number of timestamps $|t|$). As shown in Figure \ref{fig:dbave}, as the external demand amounts of microgrids over a longer period (larger $|t|$) are utilized in the $L^t$-DBSCAN algorithm, the average demand of the identified HECs becomes less (simply because less microgrids are involved in each HEC in case of larger $|t|$: as long as the external demand of each HEC exceeds $L$ at any time, the HEC cannot involve any other microgrids). Therefore, the HECs become more cohesive and the average distance between each microgrid and its centroid in the HEC would be smaller (as shown in Figure \ref{fig:dbsse}: the larger $|t|$, the smaller (spatial) sum of squared errors (SSE) \cite{WuLXCC15} of all the HECs on average).
 
Finally, similar to testing the load on transmission lines for the $K$-Means, we also compare such overall load at 2,880 timestamps for 50,000 microgrids in two cases: with HECs and without HECs. Table \ref{table:load2} also shows that the energy loss during transmission can be significantly reduced with HECs. 

\begin{table}[!h]

	\centering \caption{Load on Transmission Lines ($L^t$-DBSCAN)} 
	\begin{tabular}{|c|c|c|c|c|}
		\hline
		\small{$L$} &250000 & 550000 & 850000 & 1150000  \\
		\hline
		
		Load (with HECs) & 176127 & 329834 & 583915 & 848734 \\
		Load (without HECs) & 4951490 & 4951490 & 4951490 & 4951490\\
		\hline

	\end{tabular}
	\label{table:load2}
\end{table}

\begin{figure*}[!tbh]
	\centering 
	\subfigure[Net Energy in the MECs vs. Normalized Net Energy Distance Threshold $\epsilon$ (20,000 Mixed Microgrids with Positive Overall Net Energy)]{
		\includegraphics[angle=0, width=0.45\linewidth]{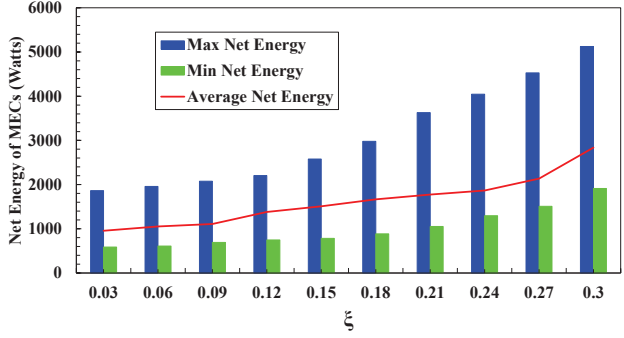}
		\label{fig:pos} } \hspace{0.1in} 
	\subfigure[Average Net Energy w.r.t. Different Number of Timestamps $|t|$ (20,000 Mixed Microgrids with Positive Overall Net Energy)]{
		\includegraphics[angle=0, width=0.45\linewidth]{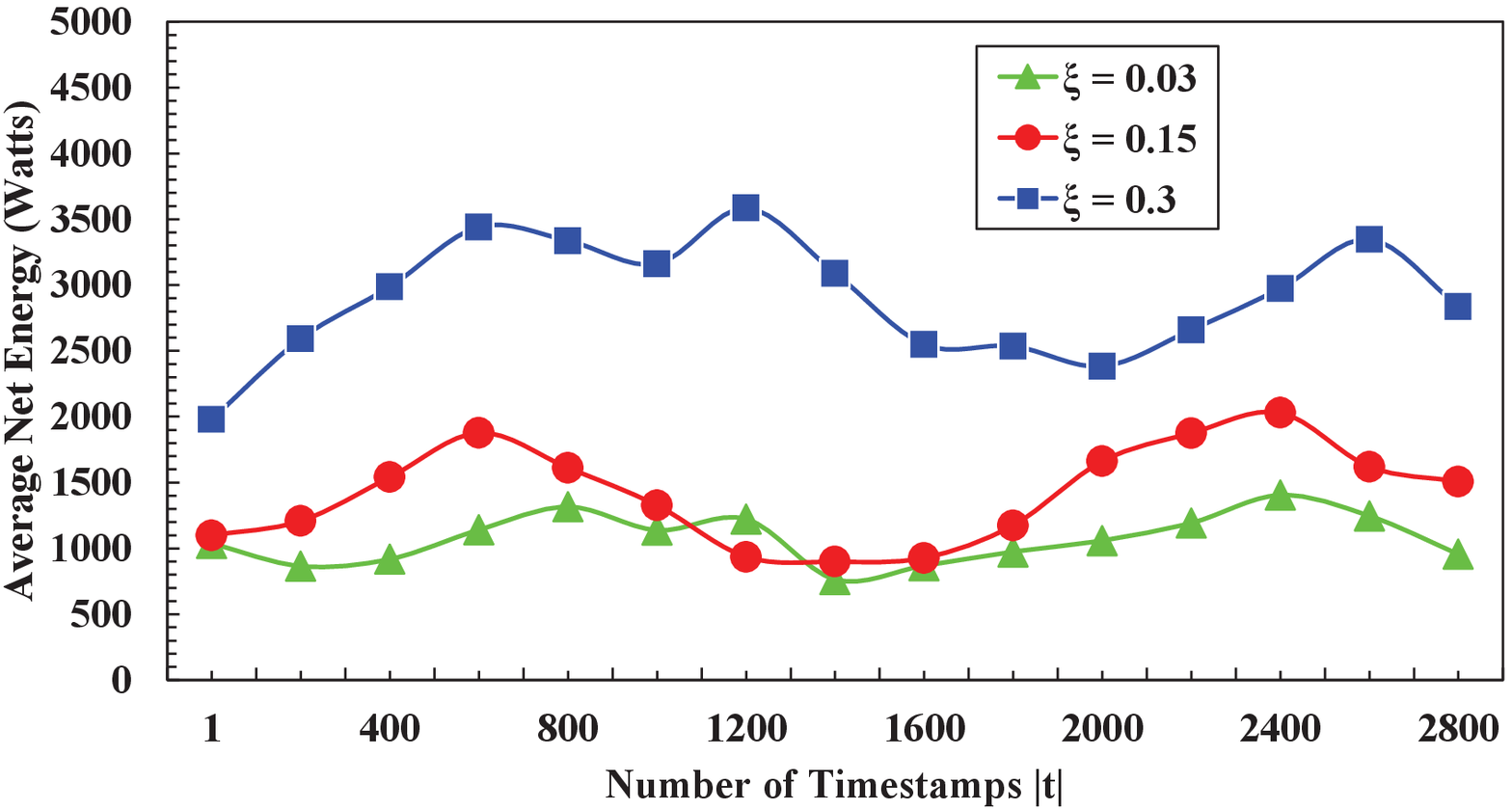}
		\label{fig:posave} } 
	\subfigure[Net Energy in the MECs vs. Normalized Net Energy Distance Threshold $\epsilon$ (20,000 Mixed Microgrids with Negative Overall Net Energy)]{
		\includegraphics[angle=0, width=0.45\linewidth]{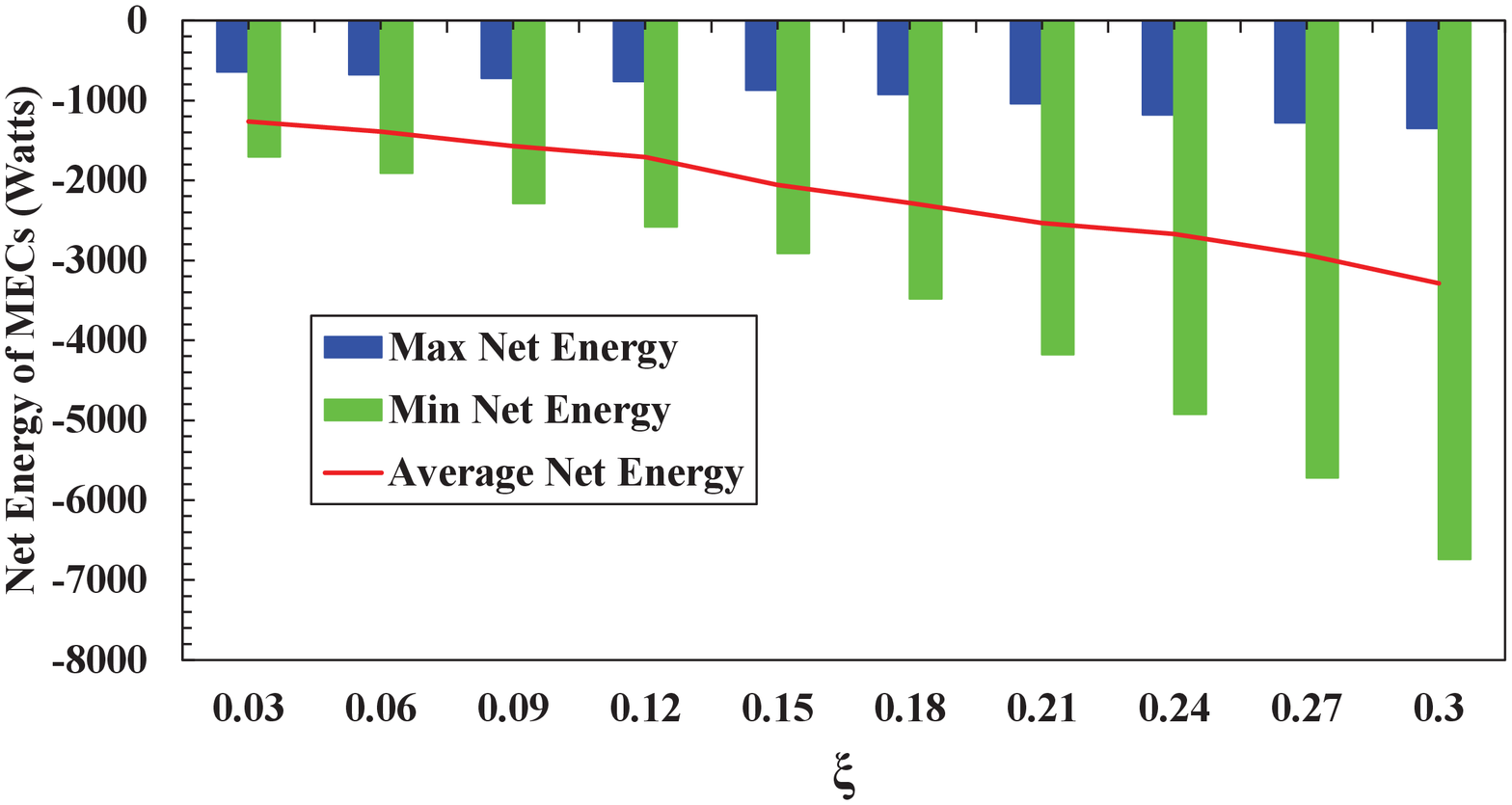}
		\label{fig:neg} } 		\hspace{0.1in} 	
	\subfigure[Average Net Energy w.r.t. Different Number of Timestamps $|t|$ (20,000 Mixed Microgrids with Negative Overall Net Energy)]{
		\includegraphics[angle=0, width=0.45\linewidth]{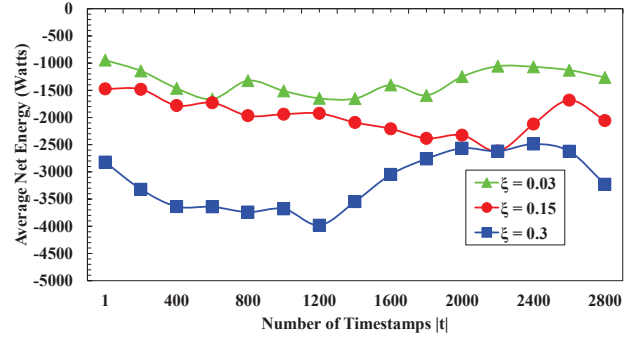}
		\label{fig:negave} } 
	\caption[Optional caption for list of figures]
	{MECs Discovery}
	\label{fig:ds}
\end{figure*}

\subsection{MECs Discovery}

The experiments for testing the MECs discovery algorithm were conducted on the synthetic dataset generated from the UMass Smart* Microgrid dataset \cite{umassdata} and the spatial dataset. Recall that the net energy of all the 50,000 microgrids (overall power generation minus overall power consumption) is \textit{negative}. To test the effectiveness of Algorithm \ref{algo:agg} in two different cases (1) positive net energy, and (2) negative net energy, we extracted two subgroups of microgrids from the 50,000 microgrids, each of which includes 20,000 microgrids and has positive and negative overall net energy at 2,880 different times, respectively. For simplicity of notations, these two subsets of microgrids are named as ``Positive'' and ``Negative'', respectively.

First, we implemented Algorithm \ref{algo:agg} with $\epsilon\in[0.03, 0.3]$ where the normalized spatial distance threshold $\epsilon'$ is fixed as a reasonable value $0.05$. Then, Figure \ref{fig:pos} shows the average, maximum and minimum net energy of all the communities generated from ``Positive'' where $\epsilon\in[0.03,0.3]$. As $\epsilon$ increases from 0.03 to 0.3, the allowed maximum differences between the overall demand and overall supply in every MEC increase significantly. The average, maximum and minimum net energy then increase as $\epsilon$ increases. Thus, the demand and supply of the MECs become better balanced with a net energy closer to 0. On the contrary, Figure \ref{fig:neg} demonstrates the results for ``Negative'', which present a reverse trend as ``Positive'', but still tend to better balanced load (net energy also becomes closer to 0) as $\epsilon$ decreases.

Second, similar to $K$-Means and $L^t$-DBSCAN, we also have some other interesting findings in the MECs discovery by utilizing microgrids' time series net energy over different lengths of periods (varying number of timestamps $|t|$). As shown in Figure \ref{fig:posave} and \ref{fig:negave}, as the net energy of microgrids over a longer period (larger $|t|$) is utilized in the MECs discovery, the average net energy of the identified MECs can have both increasing and decreasing trends. This is because larger $|t|$ can possibly lead to involving either more or less microgrids in every MEC (i.e. two microgrids' net energy distance might be large in the short term but small in the long term, and vice-versa). Then, we cannot determine whether the number of microgrids in each MEC can be increased or decreased as $|t|$ increases in Figure \ref{fig:posave} and \ref{fig:negave}. Furthermore, also in Figure \ref{fig:posave} and \ref{fig:negave}, larger $\epsilon$ would lead to a higher average net energy (positive) and lower average net energy (negative) in general. This is because larger $\epsilon$ (the threshold of net energy distance) allows more microgrids to be clustered in every MEC.

\begin{figure}[!h]
	\centering 	
	\subfigure[SSE of MECs vs. Number of Timestamps $|t|$]{
		\includegraphics[angle=0, width=0.9\linewidth]{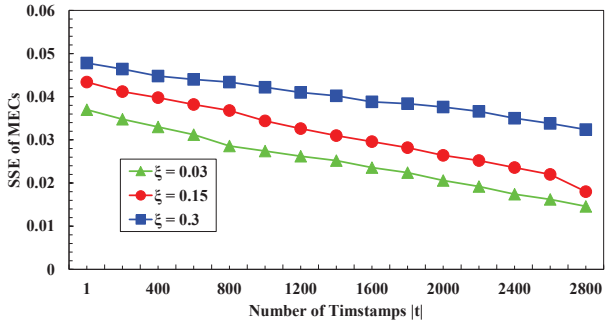}
		\label{fig:mecpsse} } \hspace{0.2in} 
	\subfigure[SSE Ratio vs. $\epsilon$]{
		\includegraphics[angle=0, width=0.9\linewidth]{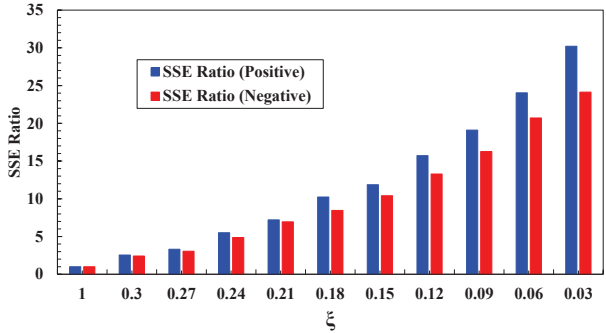}
		\label{fig:sseratio} } 
	\caption[Optional caption for list of figures]
	{(Spatial) SSE in the MECs}
	\label{fig:sse}
\end{figure}

Third, we also look at the geo-locations of the microgrids in the MECs. On one hand, we have examined the (spatial) SSE of the discovered MECs by utilizing microgrids' time series net energy over different length of periods (different $|t|$). As shown in Figure \ref{fig:mecpsse}, for any $|t|$, larger $\epsilon$ leads to higher SSE of MECs (since microgrids in the same MEC would be less cohesive if more microgrids are clustered with a larger $\epsilon$). Meanwhile, larger $|t|$ (more timestamps) results in lower SSE of MECs. This means less microgrids are clustered in each MEC as $|t|$ increases (indeed, this fact cannot be observed from Figure \ref{fig:posave} and \ref{fig:negave}). Even if larger $|t|$ gives more average number of microgrids in each MEC, since such mixed microgrids can have either positive or negative net energy, more microgrids in each MEC does not necessarily make the net energy of the MECs (positive case) higher nor make the net energy of the MECs (negative case) lower. This matches the observations in Figure \ref{fig:posave} and \ref{fig:negave}. 

On the other hand, we fixed $\epsilon=1$ and $\epsilon'=0.05$ in Algorithm \ref{algo:agg}, which then removes the constraint of NE distances and turns into a traditional agglomerative clustering problem based on geo-locations. Then, we computed the (spatial) SSE in the above case as the \emph{benchmark SSE} (say $SSE_0$) and tested how the spatial distances (viz. SSE) within each MEC vary for different levels of balanced load (different $\epsilon$). More specifically, we fix $\epsilon'=1$ (Algorithm \ref{algo:agg} only specifies the maximum NE distance threshold $\epsilon$ and removes the constraint of spatial distances), generate the MECs with $\epsilon\in[0.03, 0.3]$ for two inputs ``Positive'' and ``Negative'' respectively, and compute the corresponding (spatial) SSE for each MEC. Then, we define a new measure SSE ratio as $\frac{SSE}{SSE_0}$ and plot all the results in Figure \ref{fig:sseratio}. Clearly, the (spatial) SSE increases as $\epsilon$ declines -- an MEC with better balanced load includes the furthest microgrids from each other if the spatial distances within each MEC are not bounded (since $\epsilon'=1$). 

\begin{figure}[!h]
	\centering
	\includegraphics[angle=0, width=0.9\linewidth]{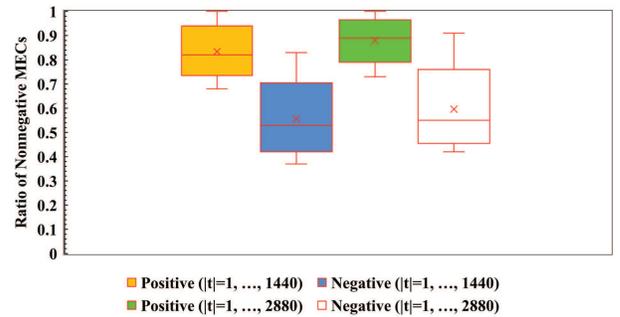} 
	\caption[Optional caption for list of figures]
	{Ratio of Nonnegative MECs (Algorithm \ref{algo:agg}: $\epsilon=0.15$, $\epsilon'=0.05$)}
	\label{fig:nonnegative}
\end{figure}

\begin{table*}[!tb]
	
	\centering \caption{SECs Discovery (Optimization-based Approach) -- 10,000 Microgrids}
	
	\begin{tabular}{|c|c|c|c|c|c|c|c|c|c|c|}
		\hline
		
		&\multicolumn{10}{|c|}{Number of Timestamps $|t|$}\\
		
		\cline{2-11}
		
		&1& 300 & 600 & 900 & 1200 & 1500 & 1800 & 2100 & 2400 & 2700\\

		\hline
		
		average net energy (all the SECs in $[T_1,T_2]$) & 732&704 &656&621&587&543&488&432&381 &324 \\
		
		\hline
		
		number of SECs: best $K$ (found by Tabu search) & 100 &110&110&120&120&130&130&130&140&140\\
		
		\hline
		
		number of micorgrids in all the SECs &10000 &10000&10000&10000&10000&10000&10000&10000&10000&10000\\
		
		\hline
		
		average number of microgrids in the SECs & 100&90.9&90.9&83.3&83.3&76.9&76.9&76.9&71.4&71.4\\
		
		\hline
		
		microgrids (with positive net energy at all times in $[T_1,T_2]$) & 6588&6588&6588&6588&6588&6588&6588&6588&6588&6588\\
		
		\hline
		
		microgrids (with negative net energy at any time in $[T_1,T_2]$) & 3412&3412&3412&3412&3412&3412&3412&3412&3412&3412\\
		
		\hline
		
		SSE (average transmission distance using SECs)  & 0.097 & 0.097 & 0.097 & 0.097 & 0.097 & 0.097 & 0.097 & 0.097 & 0.097 & 0.097\\
		\hline
		
		average distance to main grid (if no SECs) & 0.247&0.247&0.247& 0.247& 0.247& 0.247 & 0.247 & 0.247 & 0.247& 0.247\\
		\hline

	\end{tabular}
	\label{table:max}
\end{table*}

\begin{table*}[!tb]

	\centering \caption{SECs Discovery (Two-phase Algorithm) -- 10,000 Microgrids}
	
	\begin{tabular}{|c|c|c|c|c|c|c|c|c|c|c|}
		\hline
		
		&\multicolumn{10}{|c|}{Number of Timestamps $|t|$}\\
		
		\cline{2-11}
		
		&1& 300 & 600 & 900 & 1200 & 1500 & 1800 & 2100 & 2400 & 2700\\

		\hline
		
		average net energy (all the SECs in $[T_1,T_2]$) & 732 &704 &656&681&717&743&748&758&774 &789 \\
		
		\hline
		
		\hline
		
		number of SECs: best $K$ (found by $K$-Means) & 80 &80&80&80&80&80&80&80&80&80\\
		
		\hline
		
		number of micorgrids in all the SECs &10000 &10000&10000&9577&9103&8672&8557&8390&8115&8046\\
		
		\hline
		
		average number of microgrids in the SECs & 125&125&125&119.7&113.8&108.4&107.0&104.9&101.4&100.6\\
		
		\hline
		
		microgrids (with positive net energy at all times in $[T_1,T_2]$) & 6588&6588&6588&6588&6588&6588&6588&6588&6588&6588\\
		
		\hline
		
		microgrids (with negative net energy at any time in $[T_1,T_2]$) & 3412&3412&3412&3019&2515&2084&1969&1802&1527&1476\\
		
		\hline
		
		SSE (average transmission distance using SECs)  & 0.124&0.126&0.129& 0.112& 0.119& 0.118 & 0.109 & 0.106 & 0.113& 0.108\\
		\hline
		
		average distance to main grid (if no SECs) & 0.247&0.247&0.247& 0.247& 0.247& 0.247 & 0.247 & 0.247 & 0.247& 0.247\\
		\hline
		
	\end{tabular}
	\label{table:two}
\end{table*}

\begin{table*}[!tb]

	\centering \caption{SECs Discovery (Two-phase Algorithm) -- 50,000 Microgrids}
	
	\begin{tabular}{|c|c|c|c|c|c|c|c|c|c|c|}
		\hline
		
		&\multicolumn{10}{|c|}{Number of Timestamps $|t|$}\\
		
		\cline{2-11}
		
		&1& 300 & 600 & 900 & 1200 & 1500 & 1800 & 2100 & 2400 & 2700\\

		\hline
		
		average net energy (all the SECs in $[T_1,T_2]$) & 58.6 &62.3&68.7&73.5&79.4&83.5&88.7&95.3&97.3& 102 \\
		
		\hline
		
		number of SECs: best $K$ (found by $K$-Means) & 220 &220&220&220&220&220&220&220&220&220\\
		
		\hline
		
		number of micorgrids in all the SECs & 39674&38887&37981&37073&36150&35602&34821&34258&33755&33469\\
		
		\hline
		
		average number of microgrids in the SECs & 180 & 177 & 173 & 164 & 162 & 158 & 156 & 156 & 153 & 152\\
		
		\hline
		
		microgrids (with positive net energy at all times in $[T_1,T_2]$) & 25317&25317&25317&25317&25317&25317&25317&25317&25317&25317\\
		
		\hline
		
		microgrids (with negative net energy at any time in $[T_1,T_2]$) & 14357&13570&12664&11756&10833&10285&9504&8941&8438&8152\\
		
		\hline
		
		SSE (average transmission distance using SECs) & 0.142&0.147&0.14& 0.138& 0.142& 0.135 & 0.133 & 0.131 & 0.129& 0.132\\
		\hline
		
		average distance to main grid (if no SECs) & 0.223&0.223&0.223& 0.223& 0.223& 0.223 & 0.223 & 0.223 & 0.223& 0.223\\
		\hline
		
	\end{tabular}

	\label{table:two2}
\end{table*}

Furthermore, in Figure \ref{fig:nonnegative}, we present the box plot for the ratio of nonnegative MECs for two datasets (``Positive'' and ``Negative'') with $|t|\in[1,1440]$ and $|t|\in[1,2880]$, respectively (note that the MECs are identified using Algorithm \ref{algo:agg} with $\epsilon=0.15$ and $\epsilon'=0.05$). In the 1440 different results of MECs in the ``Positive'' dataset ($|t|\in[1,1440]$), the first result of MECs is obtained with $|t|=1$ (the first timestamp), the second result of MECs is obtained with $|t|=2$ (the first two timestamps), ..., the 1440th result of MECs is obtained with $|t|=1440$ (all the first 1440 timestamps). Out of the 1440 results, most of the results have more than 80\% nonnegative MECs (the highest is 100\%, and the lowest is 73.4\%). As we extend to look at all 2880 different results of MECs in the ``Positive'' dataset ($|t|\in[1,2880]$), the ratio of nonnegative MECs grows even higher. On the contrary, the ``Negative'' dataset has relatively lower ratio of nonnegative MECs (the median ratios out of 1440 results and 2880 results are 58.3\% and 61.7\%, respectively).

Finally, similar to testing the load on transmission lines for the HECs discovery, we also compare such overall load at 2,880 timestamps for 50,000 microgrids in two cases: with MECs and without MECs. Table \ref{table:load3} also shows that the energy loss during transmission can be significantly reduced with MECs.

\begin{table}[!h]

	\centering \caption{Load on Transmission Lines (MECs Discovery)} 
	\begin{tabular}{|c|c|c|c|c|}
		\hline
		\small{$\epsilon$} &0.03 & 0.12 & 0.21 & 0.3  \\
		\hline
		
		Load (with MECs) & 102795 & 291694 & 487679 & 720731 \\
		Load (without MECs) & 3265520 & 3265520  & 3265520  & 3265520 \\
		\hline
		
	\end{tabular}
	\label{table:load3}
\end{table}

\subsection{SECs Discovery}

We implemented both the optimization-based approach and the two-phase algorithm to discover the SECs. For the optimization based approach, we solved the optimization problem using the proposed Tabu Search \cite{tabu1990} based algorithm (the length of Tabu list was set as $S=10$). If the algorithm cannot find a feasible solution within 10,000 seconds, the algorithm will be terminated. As mentioned earlier, to compare the two approaches, we have generated a synthetic dataset for 10,000 microgrids with mixed net energy (more microgrids with positive net energy in $[T_1,T_2]$). Table \ref{table:max} and \ref{table:two} present the experimental results of these two approaches. We have the following observations:

\begin{itemize}
	
	\itemsep 0.2em
	
	\item Both approaches are effective to discover SECs. Optimization-based approach can assign all the microgrids to the corresponding SECs (as long as the all the constraints are satisfied). However, as a heuristic algorithm, when $|t|\geq 900$, the two-phase algorithm cannot involve all the microgrids in the SECs (feasible solution indeed exists as solved by the optimization-based approach). Among all the microgrids, the two-phase algorithm has missed some microgrids with negative net energy in $[T_1,T_2]$ as $|t|\geq 900$. Then, the average net energy of all the SECs discovered by the two-phase algorithm is greater than that of the optimization-based approach (as $|t|\geq 900$).
	
	\item The SECs discovered by the optimization-based approach are more cohesive than that of the two-phase algorithm (smaller SSE), since the optimization-based approach minimizes the SSE out of all the $K$ values. In addition, we use $K$-Means to simulate five substations of the main grid, and derive the average distance to the main grid (nearest substation) for the 10,000 microgrids, which represents the average transmission distance (from the main grid to microgrids). Then, we find out that utilizing SECs for sharing local energy can significantly reduce the energy loss in the transmission, since SSE (the average transmission distance using SECs) is far less than the average distance to main grid (0.097/0.108 vs. 0.247). Also, Table \ref{table:load4} shows that the load on transmission lines can be significantly reduced using the SECs discovered by both approaches.

\begin{table}[!h]

	\centering \caption{Load on Transmission Lines (SECs Discovery)} 
	\begin{tabular}{|c|c|c|c|c|}
		\hline
		\small{$|t|$} &1 & 600 & 1500 & 2700  \\
		\hline
		
		Load (Optimization) & 345 & 1401 & 8815 & 16765 \\
		Load (Two-phase) & 498 & 1747 & 9414 & 18609 \\
		Load (without MECs) & 2412 & 57649  & 242705 & 565504 \\
		\hline
		
	\end{tabular}
	\label{table:load4}
\end{table}	

	\item For both approaches, $K$ is selected as $\{50, 60, \dots, 200\}$, which is a reasonable set of values w.r.t. 10,000 microgrids. Table \ref{table:max} and \ref{table:two} show that the optimization-based approach identifies more SECs than the two-phase algorithm. For any $|t|$, the number of SECs identified by the two-phase algorithm is fixed (since the best $K$ is determined only by the microgrids' geo-locations with positive net energy in $[T_1,T_2]$: in the first phase). However, the optimization-based approach may identify different numbers of SECs if different $|t|$ are considered.  
	
	\item The optimization-based approach requires that all the constraints should be satisfied. Thus, such approach may not be able to find a feasible solution in some cases within a reasonable time (e.g., the 50,000 microgrids for MECs discovery). Instead, the two-phase algorithm can discover a ``maximum'' subset of microgrids to form the SECs. We also apply the two-phase algorithm to the 50,000 microgrids and present the experimental results in Table \ref{table:two2}. Similar observations can be obtained in Table \ref{table:two} and \ref{table:two2}.

	\item As discussed in Section \ref{sec:efficiency}, two-phase algorithm is more efficient than optimization-based approach (Tabu Search).

\end{itemize}

\subsection{Efficiency}
\label{sec:efficiency}

\begin{figure}[!tbh]
	\centering
	\includegraphics[angle=0, width=0.9\linewidth]{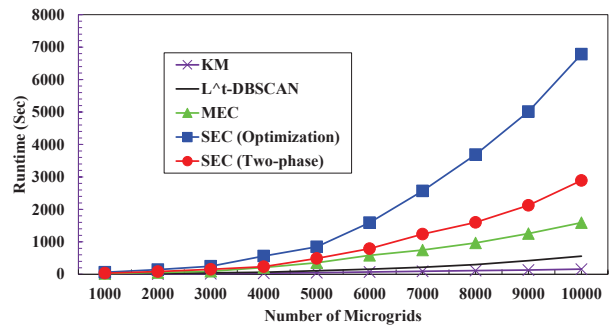} \vspace{-0.1in} 
	\caption[Optional caption for list of figures]
	{Computational Performance}
	\label{fig:runtime}
\end{figure}

Finally, we evaluated the computational performance of all the algorithms based on different input sizes (number of microgrids), and plotted the runtime of all the algorithms in Figure \ref{fig:runtime}. Note that KM, $L^t$-DBSCAN, MEC, SEC (Optimization), and SEC (Two-phase) denote five different algorithms, respectively.

Specifically, two HECs discovery algorithms ($K$-Means and $L^t$-DBSCAN) are extremely efficient with fixed parameters: the HECs number in the $K$-Means, and four parameters in the $L^t$-DBSCAN. For discovering the MECs, Algorithm \ref{algo:agg} hierarchically groups close microgrids and merges the communities based on two distance thresholds. It requires more runtime than $K$-Means and $L^t$-DBSCAN (as shown in Figure \ref{fig:runtime}). 

For SECs discovery, since both algorithms (Tabu Search and two-phase algorithm) need to find the optimal number of communities, it takes relatively longer time than HECs and MECs discovery. Indeed, in the two-phase algorithm, clustering is only applied to the microgrids with positive net energy (the first phase). Also, the step of finding microgrids with negative net energy in Algorithm \ref{algo:sscd} (the second phase) is highly efficient with a complexity of $O(N)$. Thus, the two-phase algorithm outperforms the optimization-based algorithm on runtime for the SECs discovery.

\section{Discussions}
\label{sec:discussion}

\subsection{Uncertain Energy Supply and Demand}

Recall that energy supply and demand might be stochastic on the power grid. For instance, if the power is generated using wind turbine or solar panel, the generation for power supply at the future times highly relies on the local weather and may become uncertain. Also, the energy consumption of microgrids might be uncertain since the power usage pattern of energy consumers may easily change. To address the issues of energy supply and demand uncertainty, all our energy community discovery algorithms have involved the time series energy demand and generation (net energy). Specifically,

\begin{itemize}
	\itemsep 0.2em
	
	\item For discovering a fixed number of HECs, the $K$-Means algorithm generates the communities based on the geo-locations of the microgrids, where the uncertain energy demand (or supply) does not affect the HECs.
	
	\item For discovering the HECs with bounded net energy $L$ over time $t\in[T_1,T_2]$, the $L^t$-DBSCAN algorithm ensures that the net energy of every HEC is bounded by $L$ at any time in $[T_1,T_2]$. For discovering the MECs and/or SECs, all the microgrids' net energy over period $[T_1,T_2]$ are involved for communities discovery.
	
	We can utilize a sufficiently long period $[T_1,T_2]$ in the algorithms to ensure the accuracy of the output energy communities. Note that, involving the fine-grained energy production and generation data (larger $|t|$) over a longer period $[T_1,T_2]$ in the algorithms can precisely measure the net energy of different microgrids over a sufficiently long period -- which could address the uncertainty of the net energy at future times (in a fashion of training/prediction). Since a large $|t|$ and/or a long period $[T_1,T_2]$ may result in higher computational cost, a reasonable tradeoff between the accuracy and runtime should be aware of while discovering the energy communities.
	
\end{itemize}

\subsection{Day-ahead and Real-time Energy Market}

Our energy community discovery algorithms can work effectively on both the day-ahead and real-time energy market. Specifically,

Day-ahead energy market allows the participants on the power grid to purchase and sell the electric energy at financially binding day-ahead prices for the following day (https://www.iso-ne.com). For the HECs, since such energy communities can be identified using the $K$-Means algorithm and our $L^t$-DBSCAN algorithm within a short time, the microgrids in HECs can purchase (negative net energy) or sell (positive net energy) from the main grid with the day-ahead prices for the following day. For the MECs and SECs, the algorithm takes relatively longer time to identify such energy communities, nonetheless, the MECs and SECs can be obtained within a couple of hours for even large number of microgrids (as shown in Figure \ref{fig:runtime}). In this case, all the microgrids can still purchase (negative net energy) or sell (positive net energy) from each other within every community with day-ahead prices. In summary, all the communities discovery algorithms can follow the day-ahead prices to complete their trades.
 	
 Real-time energy market allows the participants on the power grid to purchase and sell the energy using real-time prices. Even though some of our algorithms (e.g., the MECs discovery, and SECs discovery) require relatively longer time to derive the energy communities from a large number of microgrids with their time series energy generation and demand amounts, our algorithms still work effectively on the real-time energy market. The primary reason of meeting the demand of real-time market is that our algorithm can identify both short-term and long-term energy communities, by specifying different $|t|$ for energy community discovery. For larger $|t|$, the communities discovery process takes longer time but the identified HECs, MECs, and SECs can be effectively utilized for a long time. Within a longer period, all the microgrids can trade their electric energy within their communities using real-time prices (no community discovery is required in such period). For smaller $|t|$, the communities can be identified quickly, then all the microgrids can still trade energy using real-time prices.

\section{Related Work}
\label{sec:related}
As important building blocks on the smart grid, microgrids have attracted significant interests in both industry and academia in the past decade. In such context, many recent research were conducted to design microgrids and/or energy management schemes so as to improve the performance of the power grid, such as load management techniques \cite{Ahmadi2013}, demand response solutions \cite{Sarker2015}, and home automation \cite{MGOverview}. More specifically, Erol-Kantarci et al. \cite{Erol-KantarciKM11} and Dall'Anese et al. \cite{DallAneseZG13} proposed techniques for establishing microgrids on the power grid based on different criteria, such as cost minimization \cite{Erol-KantarciKM11} and power flow optimization \cite{DallAneseZG13}. In addition, analysis of data collected from distributed microgrids (e.g., demand load, energy generation and storage) has advanced the energy management of the grid and microgrids \cite{LogenthiranSS12}. Such applications include short term load forecasting for microgrids \cite{AmjadyKZ10}, load restoration for microgrids \cite{XuL11}, load shifting \cite{LiangLLLS13}, etc. 

Furthermore, some cooperative models among distributed microgrids have been investigated in multiple applications, e.g., optimizing the power loss via a unified microgrid voltage profile \cite{MaknouninejadQ14}, eliminating the central energy management unit and price coordinator via localized smart devices \cite{AsrOZC14}, load management via sharing local electricity \cite{SaadHPB12,HongIJER15,icassp18}, and load management via multiagent systems \cite{xie19}. In this paper, we develop techniques to identify communities of microgrids which can directly implement all these cooperative applications within each energy community to further advance grid performance.

In addition, large smart metering datasets collected from energy consumers have been analyzed to function many different applications \cite{HongTIFS17,energies17}. For instance, utilities can provide differentiated user services for their time-of-use energy billing plans. Pan et al. \cite{PanWH16} proposed an approach towards differentiated user services by extracting characteristic consumer load shapes from a large smart meter dataset. Diamantoulakis et al. \cite{Diamantoulakis201594} analyzed the smart metering data for load and renewable production forecasting in the electricity market and the dynamic energy management. J. Kwac and R. Rajagopal \cite{6691643} conducted big data analysis for demand response management in the smart grid infrastructure. Simmhan et al. \cite{6475927} have presented scalable machine learning models trained over big datasets for agile demand forecasting and a portal for visualizing energy consumption patterns in a cloud-based software platform.

Finally, community discovery problems generally group data objects which share similar characteristics or are close to each other, e.g., detecting communities of individuals who have similar interests on the social network \cite{YangML13}, analyzing the spatial datasets to identify geographical communities \cite{communitydiscovery}. Different from all these prior work, it might be preferable to group microgrids on the power grid which may have highly dissimilar features, e.g., two microgrids have completely opposite net energy (positive and negative). To the best of our knowledge, we take the first step to solve this new kind of community discovery problems. Note that Sanchez-Garcia et al. \cite{Fennelly2013} investigated the decomposition of power transmission network, and presented a hierarchical spectral clustering method to partition the large interconnected networks into loosely-connected zones. However, the proposed partitioning algorithm is based on the power flow optimization, whereas our community discovery problems have completely different objective functions and constraints, e.g., upper bound of the energy capacity in each community, minimizing two different distances within each community, and zero net energy.


\section{Conclusion and Future Work}
\label{sec:con}
Energy communities formed by distributed energy resources (viz. microgrids) could facilitate the power grid to advance energy management and enable microgrids to find peer microgrids to cooperate (e.g., sharing energy). In this paper, we have proposed a series of approaches to identify different energy communities for the microgrids, including homogeneous energy communities, mixed energy communities and self-sufficient energy communities. We have also validated the effectiveness and efficiency of the approaches using real world spatial dataset and power generation \& consumption datasets. 

In the future, we will investigate and solve some other variants of energy community discovery problems for microgrids. For example, the microgrids may have personalized preferences to form the energy communities (e.g., $m_1$ prefers to stay in the same community as $m_2$, rather than $m_3$), and we will try to incorporate such preferences as well as other social interactions into the energy community discovery problems. In addition, besides integrating all the energy generation and consumption over a period into the HECs, MECs, and SECs discovery, we will explore stochastic optimization models for energy community discovery based on predicting the future time series power generation and consumption, which is expected to improve the efficiency of the current energy community discovery algorithms. Finally, since the community discovery algorithms should be performed based on data collection from all the microgrids which may compromise their privacy, it is worth investigating privacy preserving models \cite{HongCikm09,OuTDSC18,meisamccs18} to analyze distributed microgrid data with limited disclosure \cite{HongJcs12,HongTDSC15}.


%



  \section*{Acknowledgments}

This research is supported in part by the National Science Foundation under the Grants No. 1745894. We also acknowledge the real world data support from the trace repository at University of Massachusetts, US and the Center for Renewable Energy Systems Technology at Loughborough University, UK.

\ifCLASSOPTIONcaptionsoff
  \newpage
\fi

\balance

\vfill



\begin{thebibliography}{10}
	
	\bibitem{uwater}
	http://www.math.uwaterloo.ca/tsp/.
	
	\bibitem{Ahmadi2013}
	M.~Ahmadi.
	\newblock {Optimizing load control for a residential microgrid in a
		collaborative environment}.
	\newblock {\em ProQuest Dissertations and Theses}, 1541222(3):66, 2013.
	
	\bibitem{AmjadyKZ10}
	N.~Amjady, F.~Keynia, and H.~Zareipour.
	\newblock Short-term load forecast of microgrids by a new bilevel prediction
	strategy.
	\newblock {\em IEEE Trans. Smart Grid}, 1(3):286--294, 2010.
	
	\bibitem{ArboleyaGCFMSBP15}
	P.~Arboleya, C.~Gonzalez{-}Moran, M.~Coto, M.~C. Falvo, L.~Martirano,
	D.~Sbordone, I.~Bertini, and B.~D. Pietra.
	\newblock Efficient energy management in smart micro-grids: {ZERO} grid impact
	buildings.
	\newblock {\em {IEEE} Trans. Smart Grid}, 6(2):1055--1063, 2015.
	
	\bibitem{AsrOZC14}
	N.~R. Asr, U.~Ojha, Z.~Zhang, and M.~Chow.
	\newblock Incremental welfare consensus algorithm for cooperative distributed
	generation/demand response in smart grid.
	\newblock {\em {IEEE} Trans. Smart Grid}, 5(6):2836--2845, 2014.
	
	\bibitem{umassdata}
	S.~Barker, A.~Mishra, D.~Irwin, E.~Cecchet, P.~Shenoy, and J.~Albrecht.
	\newblock Smart*: An open data set and tools for enabling research in
	sustainable homes.
	\newblock In {\em the 2012 Workshop on Data Mining Applications in
		Sustainability}, 2012.
	
	\bibitem{sgsg}
	S.~Goel, Y.~Hong, V.~Papakonstantinou, D.~Kloza
	\newblock Smart Grid Security.
	\newblock {\em Springer}, pages 1--129, 2015.
	
	\bibitem{ZeroEnergy}
	N.~Carlisle, O.~V. Geet, and S.~Pless.
	\newblock Definition of a "zero net energy" community.
	\newblock {\em National Renewable Energy Laboratory Technical Report},
	November, 2009.
	
	\bibitem{communitydiscovery}
	M.~Coscia, F.~Giannotti, and D.~Pedreschi.
	\newblock A classification for community discovery methods in complex networks.
	\newblock {\em Statistical Analysis and Data Mining}, 4(5):512--546, 2011.
	
	\bibitem{DallAneseZG13}
	E.~Dall'Anese, H.~Zhu, and G.~B. Giannakis.
	\newblock Distributed optimal power flow for smart microgrids.
	\newblock {\em IEEE Trans. Smart Grid}, 4(3):1464--1475, 2013.
	
	\bibitem{Diamantoulakis201594}
	P.~D. Diamantoulakis, V.~M. Kapinas, and G.~K. Karagiannidis.
	\newblock Big data analytics for dynamic energy management in smart grids.
	\newblock {\em Big Data Research}, 2(3):94 -- 101, 2015.
	
	\bibitem{Erol-KantarciKM11}
	M.~Erol-Kantarci, B.~Kantarci, and H.~T. Mouftah.
	\newblock Reliable overlay topology design for the smart microgrid network.
	\newblock {\em IEEE Network}, 25(5):38--43, 2011.
	
	\bibitem{EsterKSX96}
	M.~Ester, H.~Kriegel, J.~Sander, and X.~Xu.
	\newblock A density-based algorithm for discovering clusters in large spatial
	databases with noise.
	\newblock In {\em Proceedings of the Second International Conference on
		Knowledge Discovery and Data Mining (KDD-96), Portland, Oregon, {USA}}, pages
	226--231, 1996.
	
	\bibitem{FangMXY12}
	X.~Fang, S.~Misra, G.~Xue, and D.~Yang.
	\newblock Smart grid - the new and improved power grid: A survey.
	\newblock {\em IEEE Communications Surveys and Tutorials}, 14(4):944--980,
	2012.
	
	\bibitem{tabu1990}
	F.~Glover.
	\newblock Tabu search: A tutorial.
	\newblock {\em Interfaces}, 20(4):74--94, 1990.
	
	\bibitem{HongIJER15}
	Y.~Hong, S.~Goel, and W.~M. Liu.
	\newblock An efficient and privacy-preserving scheme for p2p energy exchange
	among smart microgrids.
	\newblock {\em International Journal of Energy Research}, 40(3):313--331, 2016.
	
	\bibitem{HongTIFS17}
	Y.~Hong, W. M.~Liu, and L.~Wang.
	\newblock  Privacy Preserving Smart Meter Streaming against Information Leakage of Appliance Status.
	\newblock {\em IEEE Trans on Information Forensics and Security}, 12(9): 2227-2241, 2017.
	
	\bibitem{sgc17}
	Y.~Hong, S.~Goel, H.~Lu, and S.~Wang.
	\newblock  Discovering Energy Communities for Microgrids on the Power Grid.
	\newblock In {\em SmartGridComm}, pages 64--70, 2017
	
	
	
	\bibitem{icassp18}
	Y.~Hong, H.~Wang, S.~Xie, and B.~Liu.
	\newblock Privacy Preserving and Collusion Resistant Energy Sharing,.
	\newblock In {\em ICASSP}, pages 6941--6945, 2018.
	
	\bibitem{HongCikm09}
	Y.~Hong, X.~He, J.~Vaidya, N.~Adam, and V.~Atluri.
	\newblock Effective Anonymization of Query Logs.
	\newblock In {\em CIKM}, pages 1465--1468, 2009.
	
	\bibitem{HongTDSC15}
	Y.~Hong, J.~Vaidya, H.~Lu, P.~Karras, and S.~Goel.
	\newblock Collaborative search log sanitization: Toward differential privacy and boosted utility.
	\newblock {\em IEEE Trans on Dependable and Secure Computing}, Vol. 12(5), pp. 504-518, 2015.
	
	\bibitem{HongJcs12}
	Y.~Hong, J.~Vaidya, and H.~Lu.
	\newblock Secure and Efficient Distributed Linear Programming.
	\newblock {\em Journal of Computer Security}, Vol. 20(5), pp. 583-634, 2012.	
	
	\bibitem{energies17}
	Y.~Hong, S.~Wang, and Z.~Huang.
	\newblock Efficient Energy Consumption Scheduling: Towards Effective Load Leveling.
	\newblock {\em Energies}, Vol. 10(1), pp. 105, 2017.	
	
	\bibitem{KanungoMNPSW02}
	T.~Kanungo, D.~M. Mount, N.~S. Netanyahu, C.~D. Piatko, R.~Silverman, and A.~Y.
	Wu.
	\newblock An efficient k-means clustering algorithm: Analysis and
	implementation.
	\newblock {\em {IEEE} Trans. Pattern Anal. Mach. Intell.}, 24(7):881--892,
	2002.
	
	\bibitem{loadmanagement}
	J.~Kennedy, P.~Ciufo, and A.~Agalgaonkar.
	\newblock Intelligent load management in microgrids.
	\newblock In {\em Power and Energy Society General Meeting, 2012 IEEE}, pages
	1--8, July 2012.
	
	\bibitem{CommunityEnergy}
	M.~King.
	\newblock Community energy: Planning, development and delivery.
	\newblock {\em International District Energy Association}, 2012.
	
	\bibitem{6691643}
	J.~Kwac and R.~Rajagopal.
	\newblock Demand response targeting using big data analytics.
	\newblock In {\em 2013 IEEE International Conference on Big Data}, pages
	683--690, Oct 2013.
	
	\bibitem{MG02}
	R.~Lasseter.
	\newblock Microgrids.
	\newblock In {\em IEEE Power Engineering Society Winter Meeting},
	volume~1, pages 305--308, 2002.
	
	\bibitem{LiangLLLS13}
	X.~Liang, X.~Li, R.~Lu, X.~Lin, and X.~Shen.
	\newblock Udp: Usage-based dynamic pricing with privacy preservation for smart
	grid.
	\newblock {\em IEEE Trans. Smart Grid}, 4(1):141--150, 2013.
	
	\bibitem{LogenthiranSS12}
	T.~Logenthiran, D.~Srinivasan, and T.~Z. Shun.
	\newblock Demand side management in smart grid using heuristic optimization.
	\newblock {\em IEEE Trans. Smart Grid}, 3(3):1244--1252, 2012.
	
		\bibitem{meisamccs18}
		M.~Mohammady, L.~Wang, Y.~Hong, H.~Louafi, M.~Pourzandi and M.~ Debbabi
		\newblock Preserving Both Privacy and Utility in Network Trace Anonymization.
		\newblock In {\em CCS}, pages 459--474, 2018
		
	\bibitem{macqueen1967}
	J.~MacQueen.
	\newblock Some methods for classification and analysis of multivariate
	observations.
	\newblock In {\em Proceedings of the Fifth Berkeley Symposium on Mathematical
		Statistics and Probability, Volume 1: Statistics}, pages 281--297, Berkeley,
	California, 1967. 
	
	\bibitem{MaknouninejadQ14}
	A.~Maknouninejad and Z.~Qu.
	\newblock Realizing unified microgrid voltage profile and loss minimization:
	{A} cooperative distributed optimization and control approach.
	\newblock {\em {IEEE} Trans. Smart Grid}, 5(4):1621--1630, 2014.
	
	
	\bibitem{OuTDSC18}
	L.~Ou, Z~Qin, S.~Liao, Y.~Hong and X.~Jia.
	\newblock  Releasing Correlated Trajectories: Towards High Utility and Optimal Differential Privacy.
	\newblock {\em IEEE Trans. on Dependable and Secure Computing}, to Appear.
		
	\bibitem{PanWH16}
	E.~Pan, D.~Wang, and Z.~Han.
	\newblock Analyzing big smart metering data towards differentiated user
	services: {A} sublinear approach.
	\newblock {\em {IEEE} Trans. Big Data}, 2(3):249--261, 2016.
	
	\bibitem{Richardson20101878}
	I.~Richardson, M.~Thomson, D.~Infield, and C.~Clifford.
	\newblock Domestic electricity use: A high-resolution energy demand model.
	\newblock {\em Energy and Buildings}, 42(10):1878 -- 1887, 2010.
	
	\bibitem{NationalGridWorkshop}
	T.~Roughan.
	\newblock Workshop on microgrid technologies and applications.
	\newblock In {\em National Grid}, 2013.
	
	\bibitem{SaadHPB12}
	W.~Saad, Z.~Han, H.~V. Poor, and T.~Basar.
	\newblock Game-theoretic methods for the smart grid: An overview of microgrid
	systems, demand-side management, and smart grid communications.
	\newblock {\em IEEE Signal Process. Mag.}, 29(5):86--105, 2012.
	
	\bibitem{Fennelly2013}
	R.~Sanchez-Garcia, M.~Fennelly, S.~Norris, N.~Wright, G.~Niblo, J.~Brodzki, and
	J.~Bialek.
	\newblock {Hierarchical Spectral Clustering of Power Grids}.
	\newblock {\em IEEE Trans. Power Systems}, 29(5):2229--2237, 2013.
	
	\bibitem{Sarker2015}
	M.~R. Sarker, M.~A. Ortega-Vazquez, and D.~S. Kirschen.
	\newblock {Optimal Coordination and Scheduling of Demand Response via Monetary
		Incentives}.
	\newblock {\em IEEE Trans. Smart Grid}, 6(3):1341--1352, 2015.
	
	\bibitem{6475927}
	Y.~Simmhan, S.~Aman, A.~Kumbhare, R.~Liu, S.~Stevens, Q.~Zhou, and V.~Prasanna.
	\newblock Cloud-based software platform for big data analytics in smart grids.
	\newblock {\em Computing in Science Engineering}, 15(4):38--47, July 2013.
	
	\bibitem{MGOverview}
	A.~Sobe and W.~Elmenreich.
	\newblock Smart microgrids: Overview and outlook.
	\newblock {\em CoRR}, abs/1304.3944, 2013.
	
	\bibitem{TanSK2005}
	P.~Tan, M.~Steinbach, and V.~Kumar.
	\newblock {\em Introduction to Data Mining}.
	\newblock Addison-Wesley, 2005.
	
	\bibitem{WuLXCC15}
	J.~Wu, H.~Liu, H.~Xiong, J.~Cao, and J.~Chen.
	\newblock K-means-based consensus clustering: {A} unified view.
	\newblock {\em {IEEE} Trans. Knowl. Data Eng.}, 27(1):155--169, 2015.
	
	\bibitem{xie19}
	S.~Xie, Y.~Hong, P.~Wan.
	\newblock A Privacy Preserving Multiagent System for Load Balancing in the Smart Grid.
	\newblock In {\em AAMAS '19}, to appear, 2019.
	
	\bibitem{XuL11}
	Y.~Xu and W.~Liu.
	\newblock Novel multiagent based load restoration algorithm for microgrids.
	\newblock {\em IEEE Trans. Smart Grid}, 2(1):152--161, 2011.
	
	\bibitem{YangML13}
	J.~Yang, J.~J. McAuley, and J.~Leskovec.
	\newblock Community detection in networks with node attributes.
	\newblock In {\em ICDM}, pages 1151--1156, 2013.
	
\end{thebibliography}
\end{document}